\documentclass[english,journal=jpca,manuscript=article,layout=twocolumn]{achemso}
\usepackage[T1]{fontenc}
\usepackage[utf8]{inputenc}
\usepackage{babel}
\usepackage{refstyle}
\usepackage{amsmath}
\usepackage{graphicx}
\usepackage{rotfloat}
\usepackage[numbers]{natbib}
\usepackage[unicode=true,pdfusetitle,
 bookmarks=true,bookmarksnumbered=false,bookmarksopen=false,
 breaklinks=false,pdfborder={0 0 1},backref=false,colorlinks=false]
 {hyperref}

\makeatletter

\title{Meta-GGA Density Functional Calculations on Atoms with Spherically
Symmetric Densities in the Finite Element Formalism}

\author{Susi Lehtola}

\email{susi.lehtola@alumni.helsinki.fi}

\affiliation{Molecular Sciences Software Institute, Blacksburg, Virginia 24061,
United States}

\alsoaffiliation{Department of Chemistry, University of Helsinki, P.O. Box 55, FI-00014
University of Helsinki, Finland}

\AtBeginDocument{\providecommand\eqref[1]{\ref{eq:#1}}}
\AtBeginDocument{\providecommand\secref[1]{\ref{sec:#1}}}
\AtBeginDocument{\providecommand\tabref[1]{\ref{tab:#1}}}
\AtBeginDocument{\providecommand\figref[1]{\ref{fig:#1}}}
\providecommand{\tabularnewline}{\\}
\RS@ifundefined{subsecref}
  {\newref{subsec}{name = \RSsectxt}}
  {}
\RS@ifundefined{thmref}
  {\def\RSthmtxt{theorem~}\newref{thm}{name = \RSthmtxt}}
  {}
\RS@ifundefined{lemref}
  {\def\RSlemtxt{lemma~}\newref{lem}{name = \RSlemtxt}}
  {}

\usepackage[version=4]{mhchem}

\SectionNumbersOn

\@ifundefined{showcaptionsetup}{}{%
 \PassOptionsToPackage{caption=false}{subfig}}
\usepackage{subfig}
\makeatother

\begin{document}
\begin{abstract}
Density functional calculations on atoms are often used for determining
accurate initial guesses as well as generating various types of pseudopotential
approximations and efficient atomic-orbital basis sets for polyatomic
calculations. To reach the best accuracy for these purposes, the atomic
calculations should employ the same density functional as the polyatomic
calculation. Atomic density functional calculations are typically
carried out employing spherically symmetric densities, corresponding
to the use of fractional orbital occupations. We have described their
implementation for density functional approximations (DFAs) belonging
to the local density approximation (LDA) and generalized gradient
approximation (GGA) levels of theory as well as Hartree--Fock (HF)
and range-separated exact exchange {[}S. Lehtola, Phys. Rev. A \textbf{2020},
\emph{101}, 012516{]}. In this work, we describe the extension to
meta-GGA functionals using the generalized Kohn--Sham scheme, in
which the energy is minimized with respect to the orbitals, which
in turn are expanded in the finite element formalism with high-order
numerical basis functions. Furnished with the new implementation,
we continue our recent work on the numerical well-behavedness of recent
meta-GGA functionals {[}S. Lehtola and M. A. L. Marques, J. Chem.
Phys. \textbf{2022}, \emph{157}, 174114{]}. We pursue complete basis
set (CBS) limit energies for recent density functionals, and find
many to be ill-behaved for the Li and Na atoms. We report basis set
truncation errors (BSTEs) of some commonly used Gaussian basis sets
for these density functionals and find the BSTEs to be strongly functional
dependent. We also discuss the importance of density thresholding
in DFAs and find that all of the functionals studied in this work
yield total energies converged to $0.1\ \mu E_{h}$ when densities
smaller than $10^{-11}a_{0}^{-3}$ are screened out.
\end{abstract}
\newcommand*\ie{{\em i.e.}}
\newcommand*\eg{{\em e.g.}}
\newcommand*\etal{{\em et al.}}
\newcommand*\citeref[1]{ref. \citenum{#1}}
\newcommand*\citerefs[1]{refs. \citenum{#1}} 

\newcommand*\Erkale{{\sc Erkale}}
\newcommand*\HelFEM{{\sc HelFEM}}
\newcommand*\Bagel{{\sc Bagel}}
\newcommand*\FHIaims{{\sc FHI-aims}}
\newcommand*\LibXC{{\sc LibXC}}
\newcommand*\Orca{{\sc Orca}}
\newcommand*\PySCF{{\sc PySCF}}
\newcommand*\PsiFour{{\sc Psi4}}
\newcommand*\Turbomole{{\sc Turbomole}}

\section{Introduction \label{sec:Introduction}}

Atoms are interesting for fundamental quantum chemistry, as they form
the simplest bound many-electron systems. The key aspect of the electronic
structure of atoms is their shell structure, which arises from the
significant amount of symmetry inherent in the Coulomb problem. Importantly
for chemistry, the shell structure of atoms is preserved to a large
extent also in polyatomic systems, because the nuclear Coulomb potential
dominates close to the nucleus, $V(r)=-Z/r\to-\infty$ when $r\to0$,
and thus the innermost electronic orbitals turn out to be insensitive
to changes in the chemical environment. This feature arguably makes
atomic calculations the keystone of electronic structure calculations:
the near-constant nature of the shell structure is the assumption
made in most computational approaches in the electronic structure
theory of polyatomic systems, as we will discuss in the following.

Quantum chemical calculations on polyatomic systems invariably start
from the solution of a self-consistent field (SCF) problem.\citep{Lehtola2020_M_1218}
The iterative solution of the SCF problem requires an initial guess
for the electron density, or the electronic orbitals. The best initial
guesses are those that correctly reproduce the shell structure of
atoms;\citep{Lehtola2019_JCTC_1593} good alternatives include the
superposition of atomic densities\citep{Almloef1982_JCC_385,VanLenthe2006_JCC_32}
(SAD) guess, the superposition of atomic potentials\citep{Lehtola2019_JCTC_1593,Lehtola2020_JCP_144105}
(SAP) guess, as well as a parameter-free extended Hückel guess\citep{Hoffmann1963_JCP_1397}
that similarly can also be derived from atomic calculations.\citep{Norman2012_CPL_229,Lehtola2019_JCTC_1593}

Also the available numerical approaches used to carry out the polyatomic
calculation are tightly connected to atomic electronic structure.
The dominant basis set for electronic structure calculations in the
solid state is plane waves; however, such calculations invariably
employ pseudopotentials\citep{Schwerdtfeger2011_C_3143} or the projector
augmented wave (PAW) method\citep{Bloechl1994_PRB_17953} that eliminate
the need for an explicit description for the chemically inactive core
electrons. Pseudopotentials and PAW setups are again derived from
atomic calculations. All-electron plane-wave calculations have only
extremely recently been shown to be feasible through the use of a
regularized nuclear Coulomb potential,\citep{Gygi2023_JCTC_1300,Lehtola2023__b}
but such calculations will likely be reserved for benchmark purposes
due to their high computational cost.

In contrast, molecular quantum chemical calculations almost invariably
employ the linear combination of atomic orbitals (LCAO) approach,
in which the $i^{\text{th}}$ molecular orbital of spin $\sigma$,
$\psi_{i\sigma}(\boldsymbol{r})$, is expressed as an expansion 
\begin{equation}
\psi_{i\sigma}(\boldsymbol{r})=\sum_{\alpha}C_{\alpha i}^{\sigma}\chi_{_{\alpha}}(\boldsymbol{r})\label{eq:LCAO}
\end{equation}
of atomic orbitals (AOs) centered at \textbf{$\boldsymbol{R}_{\alpha}$}
\begin{equation}
\chi_{\alpha}(\boldsymbol{r})=R_{n_{\alpha}l_{\alpha}}(|\boldsymbol{r}-\boldsymbol{R}_{\alpha}|)Y_{l_{\alpha}m_{\alpha}}(\widehat{\boldsymbol{r}-\boldsymbol{R}_{\alpha}}),\label{eq:AO}
\end{equation}
which are composed of radial functions $R_{nl}$ labeled by the primary
quantum number $n$ and angular quantum number $l$, combined with
spherical harmonics $Y_{lm}$ from $m=-l$ to $m=l$ in either the
real or complex form, $C_{\alpha i}^{\sigma}$ being the corresponding
expansion coefficients. Two major advantages of LCAO calculations
are that (i) all electrons can be explicitly modeled, facilitating
access to electronic core properties, for instance, and that (ii)
the basis set truncation errors turn out to be systematic in many
cases and cancel out in chemically relevant energy differences, affording
near-quantitative accuracy with basis sets of modest size.\citep{Lehtola2019_IJQC_25968}

Despite their name, the (radial) AOs used in the LCAO expansion do
not have to represent actual atomic orbitals, that is, physical one-particle
states of an actual atoms. Instead, several types of radial functions
such as Gaussian-type orbitals (GTOs) and Slater-type orbitals (STOs)
may be and are commonly used in \eqref{AO}; see \citeref{Lehtola2019_IJQC_25968}
for discussion. However, actual atomic orbitals are in principle the
best option. These AOs can be solved by a fully numerical approach,\citep{Lehtola2019_IJQC_25968}
yielding so-called numerical atomic orbitals (NAOs). NAOs are an especially
powerful basis set for electronic structure calculations: the minimal
NAO basis is already exact for non-interacting atoms in SCF calculations.\citep{Delley1990_JCP_508,Lehtola2019_IJQC_25968}
Thanks to this exactness, the issues with basis set superposition
error\citep{Boys1970_MP_553} (BSSE) that complicate the determination
of reliable molecular geometries in calculations with GTOs or STOs
are less of an issue in calculations employing NAOs, as ``borrowing''
basis functions from other atoms does not lead to improvement of atomic
energies. Combined with carefully formed basis sets,\citep{Blum2009_CPC_2175}
NAO methods have been shown to afford an excellent level of accuracy
compared to fully numerical calculations.\citep{Jensen2017_JPCL_1449}
A number of programs relying on NAOs in either all-electron or pseudopotential
form have been published and are in active use by the community.\citep{Delley2000_JCP_7756,Blum2009_CPC_2175,Larsen2009_PRB_195112,MichaudRioux2016_JCP_593,Smidstrup2019_JPCM_15901,Garcia2020_JCP_204108,Nakata2020_JCP_164112} 

Because the chemical bonding situation of an atom---and the related
deformation of the atom's electron density---is not known \emph{a
priori }in a molecule, atomic starting densities, starting potentials,
pseudopotentials and NAOs are typically computed using spherically
symmetric densities, achieved by fractional occupations of the atomic
orbitals, as this ensures that all bonding situations are described
equally, on the same footing. While such an approximation may sound
coarse compared to the behavior of the real atom, the approximation
does yield a correct shell structure, and thus offers a simple and
sensible starting point for more sophisticated calculations, such
as embedding the atom in a polyatomic calculation.

Still, to reach their best accuracy, atomic starting densities, starting
potentials, pseudopotentials and NAOs should be determined for the
same exchange-correlation functional that is employed in the polyatomic
calculation. Because different functionals lead to different orbitals,
the use of inconsistent NAOs may lead to the resurrection of issues
with BSSE, for instance, as the sought-for exactness property is not
satisfied in such a case. 

However, meta-GGA functionals are not yet supported by many atomic
solvers, especially when exact exchange is also included in the functional.\citep{Lehtola2019_IJQC_25968}
Although some programs have already been extended for meta-GGA functionals,\citep{Sun2011_PRB_35117,Yao2017_JCP_224105,Holzwarth2022_PRB_125144}
the approaches are often not fully self-consistent. For example, the
work of \citet{Sun2011_PRB_35117} appears to have used the same PAWs
for all functionals instead of optimized PAWs for each functional
including meta-GGAs, while the recent work of \citet{Doumont2022_PRB_195138}
is likewise unable to describe core electrons and generate atomic
basis functions self-consistently with meta-GGA functionals, being
limited to the GGA level for full self-consistency. Many other programs
still lack support for fully self-consistent meta-GGAs. We will show
in this work that the self-consistent implementation of meta-GGAs
for atoms with spherical symmetry is not much more demanding than
that of GGA functionals. 

A major motivation of this work are repeated queries for reliable
atomic reference data from developers of other fully numerical approaches.\bibnotemark[Gulans, Verstraete, Holzwarth]\bibnotetext[Gulans]{Andris Gulans, private communication, 2021.}\bibnotetext[Verstraete]{Matthieu Verstraete, private communication, 2021.}\bibnotetext[Holzwarth]{Natalie Holzwarth, private communication, 2021.}
Implementing any new computational method or algorithm requires being
able to test whether the new implementation is correct, and the verification
of any new atomic implementation for initial guesses, pseudopotentials
or NAOs thereby requires access to reliable, high-quality reference
data. Although the National Institute of Standards and Technology
(NIST) hosts an atomic structure database,\citep{Clark1997__} its
content is limited to calculations performed with the local density
approximation (LDA).\citep{Kotochigova1997_PRA_191,Kotochigova1997_PRA_5191}
While thorough datasets on some LDA and generalized gradient approximation
(GGA) functionals can be found in the literature,\citep{Kraisler2010_PRA_42516,Lehtola2020_PRA_12516}
sub-$\mu E_{h}$ accurate reference energies for atoms with fractional
occupations and meta-GGA functionals have not been published to this
author's best knowledge. A key goal of this work is to provide such
highly reliable reference values for use in verifying other implementations.

A further motivation of this work is the need to characterize and
study the numerical behavior of meta-GGA functionals. The present
author is a long-time developer of the Libxc library of density functionals,\citep{Lehtola2018_S_1}
which is used at present by some 40 electronic structure programs.
We have recently thoroughly examined the numerical behavior of all
the density functionals in Libxc at fixed atomic electron densities.\citep{Lehtola2022_JCP_174114}
However, as discussed in \citeref{Lehtola2022_JCP_174114}, the ultimate
test for the numerical stability of density functionals is the determination
of complete basis set (CBS) limit energies in fully numerical calculations,
as this requires accurate evaluation of the total energy in a sequence
of numerical basis sets of increasing size. Fully numerical calculations
are a demanding test of the well-behavedness of density functional
approximations (DFAs), and the ability to run such calculations on
various DFAs is a great boon for the development of novel functionals
as well as for their reliable implementation in Libxc.

For all of the above reasons, it would be appealing to be able to
run self-consistent calculations with meta-GGA functionals quickly
and reliably in extended basis sets. The finite element method (FEM)
offers an attractive solution for determining reliable NAOs and total
energies. FEM affords a variational approach to the CBS limit in fully
numerical calculations,\citep{Lehtola2019_IJQC_25968,Lehtola2019_IJQC_25945}
and atomic Hartree--Fock ground-state energies converge extremely
rapidly to the CBS limit with respect to the size of the radial basis
set when high-order numerical basis sets are used.\citep{Lehtola2019_IJQC_25945,Lehtola2023__} 

We have previously published a general atomic solver in \HelFEM{}\citep{Lehtola2018__,Lehtola2019_IJQC_25945,Lehtola2020_PRA_12516}
that is able to handle meta-GGA functionals including global hybrids
with modern finite element approaches and verified it against established
Gaussian-basis approaches in \citeref{Lehtola2019_IJQC_25945}. We
have also recently examined the role of the finite element shape functions
in meta-GGA calculations, and found that the requirements for the
numerical basis set are similar for HF as well as for density functional
calculations with LDA, GGA and meta-GGA functionals, and that Lagrange
and Hermite interpolating polynomials can either be used to pursue
the CBS limit for $\tau$-dependent meta-GGA functionals.\citep{Lehtola2023__}

However, the solver described in \citeref{Lehtola2019_IJQC_25945}
targets general wave functions, which may exhibit symmetry breaking,
while most NAO approaches assume spherically symmetric orbitals with
fractional occupations. We have recently discussed the extension of
the FEM approach to the case of spherical symmetry with LDA and GGA
orbitals as well as range-separated hybrids in \citeref{Lehtola2020_PRA_12516},
and reported non-relativistic Hartree--Fock ground states for spin-unrestricted
and spin-restricted calculations for H--Og ($1\le Z\le118$) in \citerefs{Lehtola2020_JCP_144105}
and \citenum{Lehtola2020_JCP_134108}, respectively. The extension
of the approach to meta-GGA functionals is described in this work.

The layout of this work is the following. Next, in \secref{Theory},
we will outline the key pieces of FEM and present the optimal formalism
for meta-GGA functionals with fractional occupations, reducing the
problem into separate radial subproblems for each angular momentum
$l$. Computational details including the studied selection of various
LDA, GGA, and meta-GGA functionals are presented in \secref{Computational-Details}.
Results of these functionals on the closed-shell and half-closed-shell
atoms from H to Ar are presented in \secref{Results}. We will demonstrate
that sub-$\mu E_{h}$ accurate total energies can be routinely determined
with the new code for various well-behaved meta-GGA functionals, and
that this allows the accurate determination of the truncation errors
of various Gaussian basis sets. We will also show that taking full
use of the symmetry inherent in the problem yields significant speedups
in the calculations, enabling calculations to be performed at the
CBS limit in a matter of seconds on commodity hardware. Finally, we
will examine the density thresholds employed in the various density
functionals considered in this work. The article concludes in a summary
and brief discussion in \secref{Summary-and-Discussion}. Atomic units
are used throughout, unless specified otherwise.

\section{Theory \label{sec:Theory}}

In this section, we will give a brief overview of the theory necessary
for implementing meta-GGAs with fractional occupations in non-relativistic
atomic calculations. We assume that the spin-$\sigma$ orbitals are
of the form
\begin{equation}
\psi_{\sigma nlm}(\boldsymbol{r})=R_{\sigma nl}(r)Y_{l}^{m}(\hat{\boldsymbol{r}}),\label{eq:atorb}
\end{equation}
where $R_{\sigma nl}(r)$ are the spin-$\sigma$ radial functions
for primary quantum number $n$ and angular quantum number $l$, and
$Y_{l}^{m}$ are complex-valued spherical harmonics. When each such
spin-orbital is occupied by $0\leq f_{\sigma nlm}\leq1$ electrons,
the spin-$\sigma$ electron density comes out as
\begin{equation}
n_{\sigma}(\boldsymbol{r})=\sum_{nlm}f_{\sigma nlm}\left|R_{\sigma nl}(r)Y_{l}^{m}(\hat{\boldsymbol{r}})\right|^{2}.\label{eq:eldens}
\end{equation}
To achieve a spherically symmetric spin-$\sigma$ electron density
$n_{\sigma}(r)$ that only depends on the distance to the nucleus
\begin{equation}
n_{\sigma}(r)=\sum_{nl}n_{\sigma nl}(r)=\frac{1}{4\pi}\sum_{nl}f_{\sigma nl}R_{\sigma nl}^{2}(r),\label{eq:rho-nl}
\end{equation}
one divides the total spin-$\sigma$ occupation of shell $nl$ evenly
among the $2l+1$ magnetic sublevels as $f_{\sigma nlm}=f_{\sigma nl}/(2l+1)$,
as the Unsöld theorem\citep{Unsoeld1927_AP_355}
\begin{equation}
\sum_{m=-l}^{l}Y_{l}^{m}\left(Y_{l}^{m}\right)^{*}=\frac{2l+1}{4\pi}\label{eq:unsold}
\end{equation}
leads to reduction of \eqref{eldens} to the radial-only form of \eqref{rho-nl}.

The radial orbitals are expanded in a numerical basis set $\chi_{\mu}(r)$
as

\begin{equation}
R_{\sigma nl}(r)=\sum_{\mu}C_{\mu n}^{\sigma(l)}\chi_{\mu}(r),\label{eq:orb}
\end{equation}
and the energy is minimized in terms of the orbital coefficients $C_{\mu n}^{\sigma(l)}$.
For brevity, we assume familiarity with the implementation of LDAs
and GGAs in a finite basis set approach;\citep{Lehtola2020_M_1218}
a detailed description of the procedure for atomic finite element
calculations can be found in \citerefs{Lehtola2020_PRA_12516} and
\citenum{Lehtola2019_IJQC_25945}. 

Substituting \eqref{orb} into \eqref{rho-nl} leads to the compact
expression
\begin{equation}
n_{\sigma}(r)=\sum_{\mu\nu}P_{\mu\nu}^{\sigma}\chi_{\mu}(r)\chi_{\nu}(r)\label{eq:dens}
\end{equation}
where the density matrix
\begin{equation}
\boldsymbol{P}^{\sigma}=\sum_{l}\boldsymbol{P}^{\sigma(l)}\label{eq:densmat}
\end{equation}
is a sum of density matrices arising from individual angular momenta
$l$
\begin{equation}
P_{\mu\nu}^{\sigma(l)}=\sum_{n}\frac{f_{\sigma nl}}{4\pi}C_{\mu n}^{\sigma(l)}C_{\nu n}^{\sigma(l)}.\label{eq:l-densmat}
\end{equation}
This mathematical structure was used in \citeref{Lehtola2020_PRA_12516}
to formulate an approach for LDA, GGA and hybrid functionals (including
Hartree--Fock theory) that reduces to solving a set of coupled radial
eigenvalue equations, leading to significant savings in computational
and storage requirements for the wave function. In the following,
we extend this approach to functionals that depend on the spin-$\sigma$
local kinetic energy density $\tau_{\sigma}$ and/or the density Laplacian
$\nabla^{2}n_{\sigma}$ as 
\begin{align}
E_{\text{xc}}= & \int n(\boldsymbol{r})\epsilon_{\text{xc}}(\{n_{\sigma}(\boldsymbol{r})\},\{\gamma_{\sigma\sigma'}(\boldsymbol{r})\},\nonumber \\
 & \{\tau_{\sigma}(\boldsymbol{r})\},\{\nabla^{2}n_{\sigma}(\boldsymbol{r})\}){\rm d}^{3}r\label{eq:metaGGA}
\end{align}
where $\epsilon_{\text{xc}}$ is the energy density per particle that
defines the used DFA and $\gamma_{\sigma\sigma'}=\nabla n_{\sigma}\cdot\nabla n_{\sigma'}$
is the reduced gradient. Note that DFAs (\eqref{metaGGA}) are often
also written in terms of the energy density $f_{\text{xc}}=n\epsilon_{\text{xc}}$
as
\begin{align}
E_{\text{xc}}= & \int f_{\text{xc}}(\{n_{\sigma}(\boldsymbol{r})\},\{\gamma_{\sigma\sigma'}(\boldsymbol{r})\},\nonumber \\
 & \{\tau_{\sigma}(\boldsymbol{r})\},\{\nabla^{2}n_{\sigma}(\boldsymbol{r})\}){\rm d}^{3}r.\label{eq:metaGGA-f}
\end{align}

\subsection{Kinetic Energy Density Dependent Functionals \label{subsec:Kinetic-energy-density}}

The positive-definite kinetic energy density $\tau_{\sigma}$, which
is the most popular ingredient for meta-GGAs, is given by
\begin{align}
\tau_{\sigma} & =\frac{1}{2}\sum_{nlm}f_{\sigma nlm}\nabla\psi_{\sigma nlm}^{*}(\boldsymbol{r})\cdot\nabla\psi_{\sigma nlm}(\boldsymbol{r})\label{eq:tau0}\\
 & =\sum_{nl}\tau_{\sigma nl}.\label{eq:tau}
\end{align}
Following \citet{Sala2015_PRB_35126} (see also \citerefs{Nagy1989_PRA_554}
and \citenum{Santamaria1990_JMST_35}), $\tau_{\sigma nl}$ can be
rewritten as
\begin{align}
\tau_{\sigma nl} & =\frac{|\nabla n_{\sigma nl}|^{2}}{8n_{\sigma nl}}+\frac{l(l+1)}{2}\frac{n_{\sigma nl}}{r^{2}}\label{eq:taunl-in}
\end{align}
where \eqref{rho-nl} gives 
\begin{equation}
\frac{|\nabla n_{\sigma nl}|^{2}}{8n_{\sigma nl}}=\frac{f_{\sigma nl}\left(R_{\sigma nl}'(r)\right)^{2}}{2}\label{eq:gradrhoperrho}
\end{equation}
and
\begin{align}
\tau_{\sigma nl} & =\frac{1}{2}f_{\sigma nl}\left[\left(R_{\sigma nl}'(r)\right)^{2}+l(l+1)\frac{R_{\sigma nl}^{2}(r)}{r^{2}}\right].\label{eq:taunl}
\end{align}
Substituting \eqref{orb} into \eqref{taunl} leads to our final expression

\begin{align}
\tau_{\sigma} & =\frac{1}{2}\sum_{l}P_{\mu\nu}^{\sigma(l)}\left[\chi_{\mu}'(r)\chi_{\nu}'(r)+l(l+1)\frac{\chi_{\mu}(r)\chi_{\nu}(r)}{r^{2}}\right].\label{eq:tau-final}
\end{align}
\Eqref{tau-final} is our first result: the local kinetic energy density
can be rewritten as a sum of contributions from various angular momenta,
which can be written solely in terms of radial density matrices. Furthermore,
since we use the same radial basis set for all angular momenta,\citep{Lehtola2019_IJQC_25945}
\eqref{tau-final} can be evaluated faster as 
\begin{align}
\tau_{\sigma} & =\frac{1}{2}P_{\mu\nu}^{\sigma}\chi_{\mu}'(r)\chi_{\nu}'(r)+\frac{1}{2}\overline{P}_{\mu\nu}^{\sigma}\frac{\chi_{\mu}(r)\chi_{\nu}(r)}{r^{2}}\label{eq:tau-fast}
\end{align}
where $\boldsymbol{P}^{\sigma}$ was defined in \eqref{densmat} and
we have introduced an angular-weighted density matrix
\begin{equation}
\overline{\boldsymbol{P}}^{\sigma}=\sum_{l}l(l+1)\boldsymbol{P}^{\sigma(l)}.\label{eq:Pbar}
\end{equation}

The final step needed for a SCF algorithm is the minimization of the
total energy by variation of the orbital coefficients.\citep{Lehtola2020_M_1218}
As all the DFA ingredients are now seen to be spherically symmetric,
the density functional contribution to the energy from \eqref{metaGGA-f}
simplifies to

\begin{equation}
E_{\text{xc}}=\int f_{xc}(\boldsymbol{r})d^{3}r=4\pi\int_{0}^{\infty}r^{2}f_{\text{xc}}(r){\rm d}r.\label{eq:Exc}
\end{equation}

In the LCAO approach, varying the total energy $E$ with respect to
the orbital coefficients leads to the Roothaan equation $\boldsymbol{F}^{\sigma}\boldsymbol{C}^{\sigma}=\boldsymbol{S}\boldsymbol{C}^{\sigma}\boldsymbol{E}^{\sigma}$,\citep{Lehtola2020_M_1218}
where $F_{\mu\nu}^{\sigma}=\partial E/\partial P_{\mu\nu}^{\sigma}$
is the spin-$\sigma$ Fock matrix, $\boldsymbol{C}^{\sigma}$ and
$\boldsymbol{E}^{\sigma}$ are the orbital coefficients and the corresponding
diagonal matrix of orbital energies, respectively, and \textbf{$\boldsymbol{S}$}
is the overlap matrix with elements $S_{\mu\nu}=\langle\chi_{\mu}|\chi_{\nu}\rangle$.\citep{Lehtola2020_M_1218} 

The use of spherical symmetry leads to the Roothaan equation splitting
into radial subproblems $\boldsymbol{F}^{\sigma(l)}\boldsymbol{C}^{\sigma(l)}=\boldsymbol{S}^{(l)}\boldsymbol{C}^{\sigma(l)}\boldsymbol{E}^{\sigma(l)}$
for every angular momentum $l$, where the radial Fock matrix is
\begin{equation}
\boldsymbol{F}^{\sigma(l)}=\frac{1}{4\pi}\frac{\partial E}{\partial P^{\sigma(l)}}\label{eq:angfock}
\end{equation}
and the radial overlap matrix is
\begin{equation}
S_{\mu\nu}^{(l)}=S_{\mu\nu}=\int_{0}^{\infty}r^{2}\chi_{\mu}(r)\chi_{\nu}(r){\rm d}r.\label{eq:radovl}
\end{equation}
The LDA and GGA type contributions to the radial Fock matrix are independent
of the angular momentum\citep{Lehtola2020_PRA_12516}
\begin{align}
F_{\mu\nu}^{\sigma;\text{LDA}} & =\int_{0}^{\infty}r^{2}\frac{\partial f_{\text{xc}}}{\partial n_{\sigma}}\chi_{\mu}(r)\chi_{\nu}(r){\rm d}r\label{eq:fock-LDA}\\
F_{\mu\nu}^{\sigma;\text{GGA}} & =F_{\mu\nu}^{\sigma;\text{LDA}}\nonumber \\
+ & \int_{0}^{\infty}r^{2}\left[2\frac{\partial f_{\text{xc}}}{\partial\gamma_{\sigma\sigma}}\frac{{\rm d}n_{\sigma}}{{\rm d}r}+\frac{\partial f_{\text{xc}}}{\partial\gamma_{\sigma\sigma'}}\frac{{\rm d}n_{\sigma'}}{{\rm d}r}\right]\nonumber \\
 & \times\left(\chi_{\mu}'(r)\chi_{\nu}(r)+\chi_{\mu}(r)\chi_{\nu}'(r)\right){\rm d}r,\label{eq:fock-GGA}
\end{align}
while the kinetic energy as well as exact exchange contributions to
the Fock matrix are $l$ dependent with expressions given in \citeref{Lehtola2020_PRA_12516}.
The $\tau$ dependence of meta-GGAs similarly leads to an $l$ dependent
Fock matrix contribution
\begin{align}
F_{\mu\nu}^{\sigma(l);\text{\ensuremath{\tau}-mGGA}} & =F_{\mu\nu}^{\sigma;\text{GGA}}+\frac{1}{4\pi}\frac{\partial E}{\partial P_{\mu\nu}^{\sigma(l)}}\nonumber \\
 & =F_{\mu\nu}^{\sigma;\text{GGA}}+\int_{0}^{\infty}r^{2}\frac{\partial f_{\text{xc}}}{\partial\tau_{\sigma}}\frac{\partial\tau_{\sigma}}{\partial P_{\mu\nu}^{\sigma(l)}}{\rm d}r,\label{eq:Fl}
\end{align}
which evaluates to

\begin{align}
F_{\mu\nu}^{\sigma(l);\text{\ensuremath{\tau}-mGGA}}= & F_{\mu\nu}^{\sigma;\text{GGA}}\nonumber \\
 & +\frac{1}{2}\int_{0}^{\infty}\frac{\partial f_{\text{xc}}}{\partial\tau_{\sigma}}\Bigg[r^{2}\chi_{\mu}'(r)\chi_{\nu}'(r)\nonumber \\
 & +l(l+1)\chi_{\mu}(r)\chi_{\nu}(r)\Bigg]{\rm d}r\label{eq:fock-tau}
\end{align}
which is the final piece of the implementation for $\tau$-dependent
functionals. 

\subsection{Density Laplacian Dependent Functionals \label{subsec:Density-Laplacian-dependent}}

The Laplacian of the density, $\nabla^{2}n_{\sigma}$, is straightforward
to process, as the density is spherically symmetric by construction
(\eqref{rho-nl}). One must merely remember that the Laplacian has
two terms in the spherical polar coordinate system: 
\begin{equation}
\nabla^{2}n_{\sigma}(r)=\frac{{\rm d}^{2}n_{\sigma}(r)}{{\rm d}r^{2}}+\frac{2}{r}\frac{{\rm d}n_{\sigma}(r)}{{\rm d}r}.\label{eq:dens-laplace}
\end{equation}
Substituting \eqref{dens} into \eqref{dens-laplace} yields 
\begin{align}
\nabla^{2}n_{\sigma}(r)= & \sum_{\mu\nu}P_{\mu\nu}^{\sigma}\Bigg[\frac{4\chi_{\mu}(r)\chi_{\nu}'(r)}{r}\nonumber \\
 & +2\left(\chi_{\mu}'(r)\chi_{\nu}'(r)+\chi_{\mu}(r)\chi_{\nu}''(r)\right)\Bigg]\label{eq:dens-lapl-pmat}
\end{align}

The Laplacian dependence leads to an additional contribution to the
Fock matrix given by

\begin{align}
F_{\mu\nu}^{\sigma;\text{lapl-mGGA}} & =\frac{1}{4\pi}\frac{\partial E}{\partial P_{\mu\nu}^{\sigma}}=F_{\mu\nu}^{\sigma;\text{GGA}}+\nonumber \\
 & \int_{0}^{\infty}r^{2}\frac{\partial f_{xc}}{\partial(\nabla^{2}n_{\sigma})}\frac{\partial(\nabla^{2}n_{\sigma})}{\partial P_{\mu\nu}^{\sigma}}{\rm d}r\label{eq:lapl-fock-definition}
\end{align}
which evaluates to the symmetric expression
\begin{align}
F_{\mu\nu}^{\sigma;\text{lapl-mGGA}} & =F_{\mu\nu}^{\sigma;\text{GGA}}+\nonumber \\
 & \int_{0}^{\infty}\frac{\partial f_{xc}}{\partial(\nabla^{2}n_{\sigma})}\Bigg[2r\chi_{\mu}'(r)\chi_{\nu}(r)\nonumber \\
 & +2r\chi_{\mu}(r)\chi_{\nu}'(r)+2r^{2}\chi_{\mu}'(r)\chi_{\nu}'(r)\nonumber \\
 & +r^{2}\chi_{\mu}''(r)\chi_{\nu}(r)+r^{2}\chi_{\mu}(r)\chi_{\nu}''(r)\Bigg]{\rm d}r.\label{eq:fock-lapl}
\end{align}
 Note that the expression in \eqref{fock-lapl}, like the LDA and
GGA contributions of \eqref{fock-LDA, fock-GGA}, applies to all angular
momenta, at variance to the term arising from local kinetic energy
density dependence in \eqref{fock-tau}. Note also that the first
two terms in the integral \eqref{fock-lapl} can be combined with
the GGA expression in \eqref{fock-GGA}, while the third term in \eqref{fock-lapl}
is of the same form as the first term in \eqref{fock-tau} and can
be likewise evaluated together.

\subsection{Numerical Stability Close to the Origin}

The local kinetic energy $\tau_{\sigma}$, \eqref{tau-fast}, seems
tricky to evaluate near the origin due to the second term of \eqref{taunl-in},
that is, $\sum_{nl}l(l+1)n_{\sigma nl}(r)/r^{2}$. However, it is
easy to see that the term is regular, as only $s$ orbitals have electron
density at the nucleus: the $s$ orbital contribution is killed as
$l(l+1)=0$, while $n_{\sigma nl}(r)\to0$ when $r\to0$ for $l>0$. 

A minor complication is that although $n_{\sigma nl}(r)\ge0$ by definition,
in practice $\sum_{nl}l(l+1)n_{\sigma nl}$ can attain a small negative
value due to finite numerical precision, which can be magnified by
a large $r^{-2}$ factor to generate a large negative contribution
close to the nucleus. We have found that such cases only occur in
the few quadrature points closest to the nucleus that carry small
quadrature weights. For simplicity, we opted to stabilize this term
simply by ensuring that it is non-negative by setting any negative
contributions to zero.

The density Laplacian $\nabla^{2}n_{\sigma}$, \eqref{dens-lapl-pmat},
has a clear singularity at the origin in its first term, where the
product of the numerically stable basis functions (see \eqref{radbas-stable}
below for discussion) and their derivatives is multiplied by a singular
$r^{-1}$ factor. The singularity is, however, integrable, as each
term becomes regular at the origin when multiplied by $r^{2}$ of
the quadrature weight. This also makes the Fock matrix expression
of \eqref{fock-lapl} regular, provided that $\partial f_{\text{xc}}/\partial(\nabla^{2}n_{\sigma})$
is finite when $r\to0$.

\section{Computational Details \label{sec:Computational-Details}}

\subsection{Finite Element Calculations \label{subsec:Finite-Element-Calculations}}

All finite element calculations are performed with the \HelFEM{}
program,\citep{Lehtola2018__,Lehtola2019_IJQC_25945,Lehtola2020_PRA_12516}
which employs Libxc \citep{Lehtola2018_S_1} to evaluate DFAs. The
used \HelFEM{} implementation is available in the public GitHub repository.\citep{Lehtola2018__}
The numerical basis functions used in this work are defined in terms
of piecewise polynomial finite element shape functions $B_{\mu}(r)$
as 
\begin{equation}
\chi_{\mu}(r)=\frac{B_{\mu}(r)}{r}.\label{eq:radbas}
\end{equation}
The shape functions $B_{\mu}$ in \eqref{radbas} are expressed within
each element $r\in[r_{i},r_{i+1}]$ in terms of a primitive coordinate
$x\in[-1,1]$ obtained with the transformation
\begin{equation}
x(r)=2\frac{r-r_{i}}{r_{i+1}-r_{i}}-1,\ r\in[r_{i},r_{i+1}].\label{eq:coordtrans}
\end{equation}
As in our previous works on atomic calculations,\citep{Lehtola2020_PRA_12516,Lehtola2020_JCP_144105,Lehtola2020_JCP_134108,Lehtola2019_IJQC_25945}
the shape functions are chosen to be Lagrange interpolating polynomials
(LIPs) 
\begin{equation}
B_{\mu}(x)=\prod_{\begin{array}{c}
\alpha=1\\
\alpha\neq\mu
\end{array}}^{N_{\text{nodes}}}\frac{x-x_{\alpha}}{x_{\mu}-x_{\alpha}}\label{eq:lip}
\end{equation}
with the non-uniform nodes $\{x_{\alpha}\}$ chosen from the Gauss--Lobatto
quadrature rule, which avoids the Runge instability\citep{Runge1901_ZMP_224}
and allows the use of very high-order numerical schemes. We have recently
studied the use of Hermite interpolating polynomials (HIPs) instead
of LIPs and found that HIPs and LIPs yield similar results with $\tau$-dependent
meta-GGA functionals.\citep{Lehtola2023__}

Note that although \eqref{radbas} is numerically unstable for small
$r$, we have recently shown that numerically stable basis functions
are afforded by Taylor expansions of \eqref{radbas} for small $r$:\citep{Lehtola2023__}
\begin{equation}
\chi_{\mu}(r)=\begin{cases}
r^{-1}B_{\mu}(r), & r>R\\
B_{\mu}'(0)+\frac{1}{2}B''(0)r+\dots & r\le R
\end{cases},\label{eq:radbas-stable}
\end{equation}
where $R$ is a switching radius that can be chosen automatically
by optimal matching of the left and right hand sides of the piecewise
definition in \eqref{radbas-stable}. We employ the numerically stable
form given by \eqref{radbas-stable} with high-order Taylor series
matching the polynomial order of $B_{\mu}(r)$ in all calculations
of this work, as discussed in \citeref{Lehtola2023__}.

Integrals are computed by Chebyshev quadrature with $N_{\text{quad}}$
points; a rule transformed to unit weight factor is employed for this
purpose, as it provides nodes and weights in easily computable analytical
form.\citep{PerezJorda1994_JCP_6520,Lehtola2019_IJQC_25945,Lehtola2022_JCP_174114}
All calculations discussed in this work are converged with respect
to the number of quadrature points.

The calculations employ 15-node LIPs (corresponding to $14^{\text{th}}$
order polynomials), as this order was found to be sufficient for a
rapid convergence of Hartree--Fock total energies in \citeref{Lehtola2019_IJQC_25945}.
The ``exponential grid'' of \citeref{Lehtola2019_IJQC_25945} 
\begin{equation}
r_{i}=(1+r_{\infty})^{i^{z}/N_{\text{elem}}^{z}}-1\label{eq:expgrid}
\end{equation}
is used in the present work with the default values $r_{\infty}=40a_{0}$
and $z=2$ for the practical infinity and the grid parameter,\citep{Lehtola2019_IJQC_25945}
as these values afford excellent accuracy in Hartree--Fock calculations.\citep{Lehtola2019_IJQC_25945}
We have recently shown that LDAs, GGAs, and meta-GGAs have similar
grid requirements to those of Hartree--Fock.\citep{Lehtola2023__}

The correctness of the present implementation has been verified by
direct comparison with the general implementation described in \citerefs{Lehtola2020_PRA_12516}
and \citenum{Lehtola2019_IJQC_25945}: the energy and Fock matrix
reproduced by the symmetry specialized version are in exact agreement
with those from the general implementation.

The SCF calculations are started from orbitals obtained from a tabulated
potential from a converged LDA exchange calculation,\citep{Lehtola2019_JCTC_1593,Lehtola2020_PRA_12516}
and employ a combination of Pulay's direct inversion in the iterative
subspace\citep{Pulay1982_JCC_556,Sellers1993_IJQC_31} (DIIS) and
the augmented DIIS\citep{Hu2010_JCP_54109} (ADIIS) methods for reliable
SCF convergence: we typically observe convergence within a dozen of
SCF iterations in the fully numerical basis set.

\subsection{Studied Atoms \label{subsec:Studied-Atoms}}

\begin{table*}
\begin{centering}
\begin{tabular}{lll|lll}
Atom & Term symbol & Configuration & Atom & Term symbol & Configuration\tabularnewline
\hline 
\hline 
H & $^{2}S$ & $1s$ & Ne & $^{1}S$ & $1s^{2}2s^{2}2p^{6}$\tabularnewline
He & $^{1}S$ & $1s^{2}$ & Na & $^{2}S$ & $1s^{2}2s^{2}2p^{6}3s$\tabularnewline
Li & $^{2}S$ & $1s^{2}2s$ & Mg & $^{1}S$ & $1s^{2}2s^{2}2p^{6}3s^{2}$\tabularnewline
Be & $^{1}S$ & $1s^{2}2s^{2}$ & P & $^{4}S$ & $1s^{2}2s^{2}2p^{6}3s^{2}3p^{3}$\tabularnewline
N & $^{4}S$ & $1s^{2}2s^{2}2p^{3}$ & Ar & $^{1}S$ & $1s^{2}2s^{2}2p^{6}3s^{2}3p^{6}$\tabularnewline
\end{tabular}
\par\end{centering}
\caption{Ground-state configurations used in the calculations. All calculations
feature subshells that are either filled or half-filled, corresponding
to an $S$ state ($L=0$) with a varying spin multiplicity shown by
the left-upper hand index of the term symbol. \label{tab:Configurations-used-for}}
\end{table*}

Although the presently-used implementation can handle both heavy atoms
and open shells (with the limitations of the presently examined non-relativistic
level of theory with a point nucleus and the use of fractional occupations),\citep{Lehtola2020_PRA_12516}
for simplicity, we will examine the H, He, Li, N, Ne, Na, P, and Ar
atoms at their ground-state configurations given in \tabref{Configurations-used-for}.
With the exception of H and He, the same atoms were also considered
in our recent study on the numerical ill-behavior of density functional
approximations\citep{Lehtola2022_JCP_174114} that was motivated by
this work, as many functionals were found to exhibit unsatisfactory
numerical stability in SCF calculations during the preparatory phase
of this manuscript. We have also recently examined the numerical well-behavedness
of various recent density functionals in the H atom in \citeref{Lehtola2023__}.

\subsection{Gaussian-Basis Calculations \label{subsec:Gaussian-basis-Calculations}}

Importantly, the general implementation presented in \citeref{Lehtola2019_IJQC_25945}
and the fractional-occupation version of this work coincide for the
ground states of the atoms in \tabref{Configurations-used-for} that
only feature fully occupied orbitals. This also enables the direct
comparison of the present results to those obtained with molecular
codes employing Gaussian basis sets, for example. The correctness
of the present implementation is also obvious from the excellent level
of agreement between the finite element calculations and ones performed
with benchmark-quality Gaussian basis sets\citep{Lehtola2020_JCP_134108}
that afford sub-microhartree accuracy for light elements. The Gaussian-basis
calculations in the AHGBS-9\citep{Lehtola2020_JCP_134108} and the
aug-pc-4 basis\citep{Jensen2002_JCP_3502,Jensen2002_JCP_7372,Jensen2002_JCP_9234,Jensen2003_JCP_2459,Jensen2004_JCP_3463,Jensen2007_JPCA_11198}
in fully uncontracted form (un-aug-pc-4) were performed with \Erkale{},\citep{Lehtola2012_JCC_1572,Lehtola2018__a}
which likewise employs Libxc for DFA evaluation. All basis functions
that are unnecessary to describe the ground states of \tabref{Configurations-used-for}
were removed from the Gaussian basis sets. In analogy to the finite
element calculations, the Gaussian-basis calculations were started
from error function fitted atomic LDA exchange-only potentials.\citep{Lehtola2020_JCP_144105}

\subsection{Studied Density Functionals \label{subsec:Studied-Density-Functionals}}

\begin{table*}
\begin{centering}
\begin{tabular}{llll}
Functional & Publication year & Libxc identifier & Type of functional\tabularnewline
\hline 
\hline 
HF &  &  & GH\tabularnewline
PW92\citep{Bloch1929_ZfuP_545,Dirac1930_MPCPS_376,Perdew1992_PRB_13244} & 1992 & \texttt{lda\_x+lda\_c\_pw} & LDA\tabularnewline
PBE\citep{Perdew1996_PRL_3865,Perdew1997_PRL_1396} & 1996 & \texttt{gga\_x\_pbe+gga\_c\_pbe} & GGA\tabularnewline
BLYP\citep{Becke1988_PRA_3098,Lee1988_PRB_785} & 1988 & \texttt{gga\_x\_b88+gga\_c\_lyp} & GGA\tabularnewline
B3LYP\citep{Stephens1994_JPC_11623} & 1994 & \texttt{hyb\_gga\_xc\_b3lyp} & GH GGA\tabularnewline
B97\citep{Becke1997_JCP_8554} & 1997 & \texttt{hyb\_gga\_xc\_b97} & GH GGA\tabularnewline
TPSS\citep{Tao2003_PRL_146401} & 2003 & \texttt{mgga\_x\_tpss+mgga\_c\_tpss} & meta-GGA\tabularnewline
revTPSS\citep{Perdew2009_PRL_26403,Perdew2011_PRL_179902} & 2009 & \texttt{mgga\_x\_revtpss+mgga\_c\_revtpss} & meta-GGA\tabularnewline
MS0\citep{Sun2012_JCP_51101,Perdew2009_PRL_26403} & 2012 & \texttt{mgga\_x\_ms0+gga\_c\_regtpss} & meta-GGA\tabularnewline
MVS\citep{Sun2015_PNASUSA_685,Perdew2009_PRL_26403} & 2015 & \texttt{mgga\_x\_mvs+gga\_c\_regtpss} & meta-GGA\tabularnewline
SCAN\citep{Sun2015_PNASUSA_685} & 2015 & \texttt{mgga\_x\_scan+mgga\_c\_scan} & meta-GGA\tabularnewline
rSCAN\citep{Bartok2019_JCP_161101} & 2019 & \texttt{mgga\_x\_rscan+mgga\_c\_rscan} & meta-GGA\tabularnewline
r$^{2}$SCAN\citep{Furness2020_JPCL_8208,Furness2020_JPCL_9248} & 2020 & \texttt{mgga\_x\_r2scan+mgga\_c\_r2scan} & meta-GGA\tabularnewline
r$^{2}$SCAN01\citep{Holzwarth2022_PRB_125144} & 2022 & \texttt{mgga\_x\_r2scan01+mgga\_c\_r2scan01} & meta-GGA\tabularnewline
TASKCC\citep{Aschebrock2019_PRR_33082,Schmidt2014_JCP_18} & 2019 & \texttt{mgga\_x\_task+mgga\_c\_cc} & meta-GGA\tabularnewline
$\omega$B97X-noV\citep{Mardirossian2014_PCCP_9904} & 2014 & \texttt{hyb\_gga\_xc\_wb97x\_v} & RSH GGA\tabularnewline
B97M-noV\citep{Mardirossian2015_JCP_74111} & 2015 & \texttt{mgga\_xc\_b97m\_v} & meta-GGA\tabularnewline
$\omega$B97M-noV\citep{Mardirossian2016_JCP_214110} & 2016 & \texttt{hyb\_mgga\_xc\_wb97m\_v} & RSH meta-GGA\tabularnewline
M08-HX\citep{Zhao2008_JCTC_1849} & 2008 & \texttt{hyb\_mgga\_x\_m08\_hx+mgga\_c\_m08\_hx} & GH meta-GGA\tabularnewline
MN12-SX\citep{Peverati2012_PCCP_16187} & 2012 & \texttt{hyb\_mgga\_x\_mn12\_sx+mgga\_c\_mn12\_sx} & RSH meta-GGA\tabularnewline
MN12-L\citep{Peverati2012_PCCP_13171} & 2012 & \texttt{mgga\_x\_mn12\_l+mgga\_c\_mn12\_l} & meta-GGA\tabularnewline
MN15\citep{Yu2016_CS_5032} & 2016 & \texttt{hyb\_mgga\_x\_mn15+mgga\_c\_mn15} & GH meta-GGA\tabularnewline
MN15-L\citep{Yu2016_JCTC_1280} & 2016 & \texttt{mgga\_x\_mn15\_l+mgga\_c\_mn15\_l} & meta-GGA\tabularnewline
revM06\citep{Wang2018_PNASUSA_10257} & 2018 & \texttt{hyb\_mgga\_x\_revm06+mgga\_c\_revm06} & GH meta-GGA\tabularnewline
revM06-L\citep{Wang2017_PNASUSA_8487} & 2017 & \texttt{mgga\_x\_revm06\_l+mgga\_c\_revm06\_l} & meta-GGA\tabularnewline
M06-SX\citep{Wang2020_PNASUSA_2294} & 2020 & \texttt{hyb\_mgga\_x\_m06\_sx+mgga\_c\_m06\_sx} & RSH meta-GGA\tabularnewline
revM11\citep{Verma2019_JPCA_2966} & 2019 & \texttt{hyb\_mgga\_x\_revm11+mgga\_c\_revm11} & RSH meta-GGA\tabularnewline
\end{tabular}
\par\end{centering}
\caption{Density functionals studied in this work including the relevant literature
references, accompanied by the publication year and the used Libxc
identifier. The last column identifies the type of the functional:
(semi)local LDA, GGA, or meta-GGA versus global hybrids (GHs) versus
range-separated hybrids (RSHs). HF is a GH with 100\% exact exchange
and no semilocal energy term. \label{tab:Density-functionals-studied}}
\end{table*}

Guided by the exploratory calculations and the work presented in \citeref{Lehtola2022_JCP_174114},
the density functionals considered in this work along with their literature
references are shown in \tabref{Density-functionals-studied}. 

The baseline of the selection is formed by HF,\bibnote{To simplify the discussion, we refer to Hartree--Fock as a density functional  that lacks a semilocal energy expression of the form of \eqref{metaGGA-f} and that instead relies on 100\% exact exchange.}
the 1992 Perdew--Wang (PW92) LDA, the Perdew--Burke--Ernzerhof
(PBE) and Becke--Lee--Yang--Parr (BLYP) GGAs, the B3LYP and B97
global hybrid GGAs, as well as the $\omega$B97X-V range-separated
hybrid GGA without non-local correlation ($\omega$B97X-noV). This
baseline comprised of 7 functionals is compared to 24 meta-GGAs that
consist of 15 semilocal meta-GGAs, 5 global hybrid meta-GGAs, and
4 range-separated hybrid meta-GGAs.

The meta-GGAs include the Tao--Perdew--Scuseria--Staroverov (TPSS)
meta-GGA as well as its revision (revTPSS), both of whose correlation
parts we have recently found to be numerically ill-behaved for alkali
atoms at fixed electron density.\citep{Lehtola2022_JCP_174114} Next,
the MS0, MVS, and SCAN functionals of Perdew and coworkers were included
since they have been found to exhibit successively degenerating numerical
behavior.\citep{Lehtola2022_JCP_174114} 

The SCAN functional has already been found to be ill-behaved in fully
numerical calculations by \citeauthor{Bartok2019_JCP_161101}, who
suggested the rSCAN functional where the ill behavior is fixed by
a well-behaved polynomial expansion.\citep{Bartok2019_JCP_161101}
rSCAN was then later used by \citet{Furness2020_JPCL_8208} to build
the r$^{2}$SCAN functional. r$^{2}$SCAN has showed extremely promising
accuracy in applications;\citep{Grimme2021_JCP_64103,Ehlert2021_JCP_61101}
however, it has been found to be ill-behaved in fully numerical calculations
by \citet{Holzwarth2022_PRB_125144}, who proposed another modification,
the r$^{2}$SCAN01 functional. (Note that we found the whole SCAN
family to be ill-behaved in \citeref{Lehtola2022_JCP_174114}.) The
TASKCC functional recommended by \citet{Lebeda2022_PRR_23061} is
included as another recent meta-GGA, which appeared to be well-behaved
in our recent studies.\citep{Lehtola2022_JCP_174114,Lehtola2023__}

All the meta-GGA functionals examined in this work depend only on
$\tau$. The reason for this is that Laplacian dependence is mainly
included only in older meta-GGA functionals, which are well-known
to be ill-behaved. Excluding such functionals leaves only functionals
that depend on $\tau$. Deorbitalized functionals,\citep{MejiaRodriguez2017_PRA_52512}
which replace the $\tau$ dependence in recent DFAs with the density
Laplacian through the use of a Laplacian-dependent kinetic energy
functional are an exception to this rule; however, we have found many
such functionals to be numerically ill-behaved already at fixed electron
densities.\citep{Lehtola2022_JCP_174114} Moreover, it is well-known
that kinetic energy functionals have singularities near the nucleus
that pose issues for numerics and the stability of SCF calculations,\citep{Karasiev2012_CPC_2519}
which are an issue even when pseudopotentials are used.\citep{Xia2015_PRB_45124} 

The basis function requirements for $\tau$-dependent functionals
were investigated in depth in \citeref{Lehtola2023__}, where it was
found that the LIP basis reproduces the correct CBS limit even though
it does not explicitly guarantee a continous $\tau$ by construction
unlike the Hermite interpolating polynomial (HIP) basis investigated
in \citeref{Lehtola2023__}. We tentatively attributed this success
to the action of the variational theorem: as discussed in \citeref{Lehtola2019_IJQC_25945},
discontinuous derivatives would lead to higher kinetic energies, which
are disincentivized by the variational minimization of the total energy. 

Note that calculations on Laplacian dependent functionals---which
are not considered in this work---should use at least a second-order
HIP basis set in order to make $\nabla^{2}n$ well-defined everywhere.
Such machinery was presented in \citeref{Lehtola2023__}, to which
we refer for further details.

\section{Results \label{sec:Results}}

\subsection{Convergence to the Basis Set and Density Threshold Limit \label{subsec:Convergence-to-the}}

We aim for total energies accurate to $0.1\mu E_{h}$ with respect
to all parameteres in the present calculations. In practice, we determine
that two calculations have converged to within $0.1\mu E_{h}$ precision
if the energies $E_{1}$ and $E_{2}$ of the two calculations agree
within $0.04\mu E_{h}$: $|E_{1}-E_{2}|<4\times10^{-8}$.

We therefore carry out a large number of calculations to determine
reference total energies converged with respect to all the parameters
controlling the calculation: in addition to the number of radial basis
functions, we also study the convergence with respect to the number
of quadrature points, as well as the effect of the density screening
threshold employed in the DFA implementation.

When the values of the density functional and its derivatives are
evaluated on the numerical quadrature grid by Libxc,\citep{Lehtola2018_S_1}
points with insignificant electron density $n(\boldsymbol{r})\le\epsilon$
as defined by a preset threshold $\epsilon>0$ are skipped, setting
the energy density $f_{\text{xc}}$ and all its derivatives to zero.
The rationale for such thresholding is grounded on the physical observation
that both factors in the energy density $f_{\text{xc}}=n\epsilon_{\text{xc}}(n,\dots)$
decrease when $n\to0$ and therefore points with negligible density
do not contribute meaningfully to the total energy; a practical issue
is also that points with extremely small densities often yield divergent
derivatives. Because of this, such thresholding is commonly used in
density functional implementations. 

This also extends further: typical density functional quadrature approaches
discard basis functions with negligible values. For instance, the
approach of \citet{Stratmann1996_CPL_213} discards atomic basis functions
$\chi_{\alpha}(\boldsymbol{r})$ in the sphere around the atom where
$|\chi_{\alpha}(\boldsymbol{r})|<\tilde{\epsilon}$ with $\tilde{\epsilon}=10^{-10}$;
such an approach also leads to errors in the electron density, and
the approach thereby relies on the small errors in the density not
having a significant effect on the total energy. 

A threshold $\epsilon=10^{-12}$ has been used in previous studies
with \HelFEM{}, which is also the default in the Psi4\citep{Smith2020_JCP_184108}
and ORCA\citep{Neese2020_JCP_224108} programs, for example. In this
work, we determine the density cutoff $\epsilon$ used in the evaluation
of the density functional, specified with the Libxc function \texttt{xc\_func\_set\_dens\_threshold},
by considering a series of calculations performed with decreasing
values in $\epsilon=10^{-n}$ with $n=8,\dots,15$. (The default value
used in Libxc for most density functionals for three-dimensional systems
is $\epsilon=10^{-15}$, below which value numerical issues are often
encountered due to the use of double precision arithmetic.) We deem
convergence to have been reached at the largest value of $\epsilon$
for which the energy does not change any more. 

We wish to underline here that the use of density thresholds in this
work is not an approximation that is special to this work, as such
finite thresholds are always used to ensure that the numerical implementation
of any functional is stable. Instead, the density threshold should
be considered part of the definition of the DFA, and a threshold that
is small enough to reproduce the converged value should always be
used. Unfortunately, the issue is that the values of the used thresholds
are typically not discussed in articles suggesting novel DFAs. However,
we do find that the studied functionals converge quickly in the density
threshold.

We determine the convergence to the CBS limit by considering calculations
with a sequence of increasing numerical basis sets composed of 5,
10, 15, 20, 25, 30, and 35 radial 15-node LIP elements with with 69,
139, 209, 279, 349, 419, and 489 radial basis functions, respectively.
Convergence to the CBS limit is established when the difference in
total energy to the calculation with the largest number of radial
basis functions is below the wanted precision. The convergence with
respect to the quadrature is checked by doubling the used number of
quadrature points, and checking whether the total energies of the
SCF calculations performed with different quadratures agree.

All fully numerical calculations failed with the MVS and SCAN functionals,
and these functionals were excluded from all analyses; the ill-behavedness
of SCAN was already reported by \citet{Bartok2019_JCP_161101}. The
obtained fully numerical reference energies for the remaining functionals
are shown where available in \tabref{H-N} for H, He, Li, Be, and
N and in \tabref{Ne-Ar} for Ne, Na, Mg, P, and Ar. Our analysis in
\tabref{H-N,Ne-Ar} also distinguishes cases where reliable reference
energies were not achieved by the present computational scheme due
to (i) failures with SCF convergence, (ii) failures with achieving
the CBS limit, and (iii) floating point errors in the calculation. 

As expected from our previous experience, Li, Be, Na, and Mg are a
challenge for many functionals due to their loosely bound outer electrons,
as can be seen from the large number of failed calculations for these
atoms. Although the initial guess is good, as suggested by DIIS errors
in the range of $10^{-2}$, we observe that many of the failed calculations
go wrong already in the first iteration. For instance, the revM06,
revM06-L, M06-SX, MN12-L, MN12-SX, MN15, B97M-V, and $\omega$B97M-V
calculations on Li start from a sensible total energy from the initial
guess, but jump up thousands of Hartrees in energy upon the diagonalization
of the first Fock matrix, which suggests that there are issues in
these functionals' numerical behavior for some densities.

These observations motivate a systematical examination of the initial
guess. The largest DIIS errors and largest energy changes in the first
iteration that arise from the first full diagonalization of the Fock
matrix are shown in \tabref{Largest-initial-DIIS}. The functionals
that stand out with a large DIIS error are MS0 and MVS, and all nine
Minnesota functionals. The Ar atom has the largest initial DIIS error
out of the studied atoms and functionals, with the exception of PW92
that encounters its largest initial DIIS error for the N atom. 

Interestingly, the initial DIIS error does not appear to correlate
strongly with the initial change in energy. Only B97, the Berkeley
meta-GGA functionals B97M-noV and $\omega$B97M-noV, and six out of
nine Minnesota functionals show an increase in energy upon the first
diagonalization; however, these increases are alarmingly large. We
note again that the largest stability problems appear to be encountered
with the Li and Na atoms, as can be seen from \tabref{Largest-initial-DIIS}. 

\begin{table*}
\begin{centering}
\begin{tabular}{l|lr|lr}
Method & Atom & $ \epsilon_\text{DIIS} $ & Atom & $ \Delta E $ \\
\hline
\hline
HF & Ar & $ 5.4\times 10^{-2}$ & He & $-2.8\times 10^{-3}$ \\
PW92 & N & $ 2.4\times 10^{-3}$ & He & $-2.2\times 10^{-4}$ \\
PBE & Ar & $ 4.9\times 10^{-2}$ & He & $-1.6\times 10^{-3}$ \\
BLYP & Ar & $ 5.4\times 10^{-2}$ & He & $-2.0\times 10^{-3}$ \\
B3LYP & Ar & $ 3.9\times 10^{-2}$ & He & $-1.9\times 10^{-3}$ \\
B97 & Ar & $ 4.2\times 10^{-2}$ & Li & $ 7.2\times 10^{0}$ \\
TPSS & Ar & $ 5.0\times 10^{-2}$ & He & $-2.6\times 10^{-3}$ \\
revTPSS & Ar & $ 5.6\times 10^{-2}$ & He & $-2.6\times 10^{-3}$ \\
MS0 & Ar & $ 2.0\times 10^{-1}$ & He & $-2.4\times 10^{-3}$ \\
MVS & Ar & $ 4.2\times 10^{-1}$ & He & $-2.4\times 10^{-3}$ \\
rSCAN & Ar & $ 5.1\times 10^{-2}$ & He & $-2.2\times 10^{-3}$ \\
r$^2$SCAN & Ar & $ 5.5\times 10^{-2}$ & He & $-2.2\times 10^{-3}$ \\
r$^2$SCAN01 & Ar & $ 5.5\times 10^{-2}$ & He & $-2.2\times 10^{-3}$ \\
TASKCC & Ar & $ 8.4\times 10^{-2}$ & He & $-3.5\times 10^{-3}$ \\
$\omega$B97X-noV & Ar & $ 4.8\times 10^{-2}$ & He & $-1.8\times 10^{-3}$ \\
B97M-noV & Ar & $ 7.3\times 10^{-2}$ & Li & $ 5.7\times 10^{4}$ \\
$\omega$B97M-noV & Ar & $ 8.0\times 10^{-2}$ & Li & $ 5.7\times 10^{4}$ \\
M08-HX & Ar & $ 4.9\times 10^{-1}$ & H & $-4.3\times 10^{-3}$ \\
MN12-SX & Ar & $ 1.2\times 10^{0}$ & Na & $ 1.1\times 10^{5}$ \\
MN12-L & Ar & $ 5.8\times 10^{-1}$ & Li & $ 7.6\times 10^{4}$ \\
MN15 & Ar & $ 1.5\times 10^{0}$ & Li & $ 1.1\times 10^{5}$ \\
MN15-L & Ar & $ 1.5\times 10^{0}$ & H & $-4.0\times 10^{-3}$ \\
revM06 & Ar & $ 1.5\times 10^{-1}$ & Li & $ 3.1\times 10^{3}$ \\
revM06-L & Ar & $ 2.5\times 10^{-1}$ & Na & $ 5.4\times 10^{3}$ \\
M06-SX & Ar & $ 1.5\times 10^{-1}$ & Li & $ 3.1\times 10^{3}$ \\
revM11 & Ar & $ 2.6\times 10^{-1}$ & He & $-2.7\times 10^{-3}$ \\
\end{tabular}

\par\end{centering}
\caption{Largest initial DIIS error $\epsilon_{\text{DIIS}}$ and the largest
energy change in the first iteration in the calculations with 35 radial
elements and a density threshold $\epsilon=10^{-12}$. The corresponding
atoms are also shown. \label{tab:Largest-initial-DIIS}}

\end{table*}

Our baseline of HF, PW92, PBE, BLYP, B3LYP, and B97 converge without
issues, with the exceptions of Li and N with B97. The B97 functional
appears to be less smooth than the other baseline functionals, as
evidenced by its need for more radial elements to reach the same convergence.

The TPSS, revTPSS and TASKCC functionals are well-behaved, easily
converging to the CBS limit for all atoms with the exception of Li,
which required surprisingly many elements.

The MS0 functional is ill-behaved, failing to reach the CBS limit
for Li, Be, Na, Mg, and P even with the extended numerical basis sets
considered in this work.

The rSCAN functional is well-behaved, other than failing to reach
the CBS limit for Li. In partial agreement with \citet{Holzwarth2022_PRB_125144},
we find that the r$^{2}$SCAN01 functional is better-behaved than
r$^{2}$SCAN, as calculations failed for Li and Na for the latter
functional. However, we do not find evidence that r$^{2}$SCAN01 is
otherwise smoother, as the functional still takes more radial elements
to converge than the well-behaved TPSS, revTPSS or TASKCC functionals,
and as r$^{2}$SCAN01 required more radial elements to converge P
than r$^{2}$SCAN did.

The $\omega$B97X-noV range-separated hybrid GGA is well-behaved and
converges easily for all studied systems. In contrast, the B97M-noV
and $\omega$B97M-noV meta-GGAs fail to converge for Li, N, Na, and
P. All of the failed calculations are characterized by large jumps
in energy in the first iteration, as discussed above for several functionals
with Li and Na.

The nine Minnesota functionals appear well-behaved for H, He, N, Ne,
and Ar, as all nine functionals reach converged CBS limit energies.
The M08-HX functional, however, requires many more radial elements
than the other functionals, which can be explained by our recent observation
in \citeref{Lehtola2023__} of sharp non-physical behavior in the
functional. No Minnesota functional is successful for Li, while only
revM11 is able to reach a converged CBS limit for Na. Only MN15, revM06,
revM06-L, and M06-SX reach CBS limits for Be. For Mg, the CBS limit
is reached by MN15, MN15-L, revM06-L, and revM11. All the studied
Minnesota functionals except MN12-SX reach a CBS limit for P.

The converged density thresholds are shown in \tabref{Density-thresholds};
the systems that failed to converge in \tabref{H-N} were excluded
in this analysis. All functionals reach total energies converged to
$0.1\mu E_{h}$ with a density threshold of $\epsilon=10^{-11}$ or
larger, confirming the general reliability of the universally used
screening approach, and confirming the reliability of the results
previously obtained with the default threshold $\epsilon=10^{-12}$.

\begin{table*}
\small
\begin{tabular}{llllll}
Functional & H & He & Li & Be & N\\
\hline\hline
HF & $ -0.5000000 $/5 & $ -2.8616800 $/5 & $ -7.4327509 $/5 & $ -14.5730232 $/5 & $ -54.4045483 $/5\\
PW92 & $ -0.4787107 $/5 & $ -2.8344552 $/5 & $ -7.3432842 $/5 & $ -14.4464735 $/5 & $ -54.1343867 $/5\\
PBE & $ -0.4999904 $/5 & $ -2.8929349 $/5 & $ -7.4621804 $/10 & $ -14.6299477 $/10 & $ -54.5357555 $/5\\
BLYP & $ -0.4979143 $/5 & $ -2.9070669 $/5 & $ -7.4826660 $/5 & $ -14.6615080 $/5 & $ -54.5931773 $/5\\
B3LYP & $ -0.5024433 $/5 & $ -2.9152187 $/5 & $ -7.4929571 $/5 & $ -14.6733282 $/5 & $ -54.6070284 $/5\\
B97 & $ -0.5029846 $/5 & $ -2.9099945 $/5 & NoSCF & $ -14.6671376 $/10 & NoSCF\\
\hline
TPSS & $ -0.5002355 $/5 & $ -2.9096639 $/5 & $ -7.4891131 $/15 & $ -14.6717170 $/10 & $ -54.6161733 $/10\\
revTPSS & $ -0.5001577 $/5 & $ -2.9120536 $/5 & $ -7.4901709 $/20 & $ -14.6725883 $/10 & $ -54.5978896 $/10\\
TASKCC & $ -0.5001730 $/5 & $ -2.9794485 $/5 & $ -7.5620745 $/15 & $ -14.7471258 $/10 & $ -54.6252881 $/10\\
MS0 & $ -0.5066733 $/5 & $ -2.9115322 $/5 & NoCBS & NoCBS & $ -54.6179098 $/20\\
rSCAN & $ -0.5001732 $/5 & $ -2.9049561 $/5 & NoCBS & $ -14.6511980 $/10 & $ -54.5993803 $/20\\
r$^2$SCAN & $ -0.5001732 $/5 & $ -2.9049561 $/5 & NoSCF & $ -14.6490866 $/15 & $ -54.5840337 $/20\\
r$^2$SCAN01 & $ -0.5001732 $/5 & $ -2.9049561 $/5 & $ -7.4800036 $/20 & $ -14.6496022 $/15 & $ -54.5860431 $/20\\
\hline
$\omega$B97X-noV & $ -0.5053272 $/5 & $ -2.9127069 $/5 & $ -7.4924999 $/5 & $ -14.6805446 $/5 & $ -54.6197329 $/5\\
B97M-noV & $ -0.5061077 $/5 & $ -2.9367807 $/5 & NoSCF & $ -14.7081142 $/10 & NoSCF\\
$\omega$B97M-noV & $ -0.4992064 $/5 & $ -2.9080420 $/5 & NoSCF & $ -14.6807256 $/10 & NoSCF\\
\hline
M08-HX & $ -0.5039981 $/10 & $ -2.9181530 $/20 & NoSCF & NoCBS & $ -54.5963110 $/20\\
MN12-SX & $ -0.4970768 $/5 & $ -2.9170941 $/10 & NoSCF & NoSCF & $ -54.5845822 $/15\\
MN12-L & $ -0.4923232 $/5 & $ -2.9156167 $/10 & NoSCF & NoCBS & $ -54.5673401 $/10\\
MN15 & $ -0.4997453 $/5 & $ -2.9219234 $/5 & NoSCF & $ -14.6808919 $/15 & $ -54.5889705 $/10\\
MN15-L & $ -0.4965988 $/5 & $ -2.9161651 $/5 & NoSCF & NoCBS & $ -54.5963736 $/10\\
revM06 & $ -0.4978698 $/5 & $ -2.9129975 $/5 & NoSCF & $ -14.6643064 $/15 & $ -54.5822636 $/10\\
revM06-L & $ -0.5000720 $/5 & $ -2.9239856 $/5 & NoSCF & $ -14.6738826 $/20 & $ -54.5927481 $/10\\
M06-SX & $ -0.4856891 $/5 & $ -2.8992554 $/5 & NoSCF & $ -14.6468389 $/10 & $ -54.5676518 $/10\\
revM11 & $ -0.5023467 $/5 & $ -2.8975261 $/5 & NoCBS & NoCBS & $ -54.5778168 $/10\\
\end{tabular}

\caption{Reference energies in $E_{h}$ with the finite element method / number
of radial elements required to reach the converged energy for the
H, He, Li, Be, and N atoms. Calculations for which reliable reference
energies marked as follows. NoSCF: the SCF procedure failed to converge.
NoQuad: the numerical quadrature was not converged. NoCBS: a converged
energy was not reached. NaN: a not-a-number floating point error was
encountered in the calculations \label{tab:H-N}}
\end{table*}

\begin{table*}
\small
\begin{tabular}{llllll}
Functional & Ne & Na & Mg & P & Ar\\
\hline\hline
HF & $ -128.5470981 $/5 & $ -161.8589538 $/5 & $ -199.6146364 $/5 & $ -340.7192753 $/5 & $ -526.8175128 $/5\\
PW92 & $ -128.2299172 $/5 & $ -161.4436320 $/5 & $ -199.1352883 $/5 & $ -340.0000523 $/5 & $ -525.9397934 $/5\\
PBE & $ -128.8664277 $/5 & $ -162.1726872 $/5 & $ -199.9551151 $/5 & $ -341.1156817 $/10 & $ -527.3461288 $/5\\
BLYP & $ -128.9730149 $/5 & $ -162.2927034 $/5 & $ -200.0926430 $/5 & $ -341.2778807 $/5 & $ -527.5510394 $/5\\
B3LYP & $ -128.9809732 $/5 & $ -162.3031506 $/5 & $ -200.1035499 $/5 & $ -341.2928849 $/5 & $ -527.5678350 $/5\\
B97 & $ -128.9418808 $/10 & $ -162.2557399 $/10 & $ -200.0507705 $/10 & $ -341.2270703 $/10 & $ -527.4847536 $/10\\
\hline
TPSS & $ -128.9811078 $/10 & $ -162.2986086 $/10 & $ -200.0927812 $/10 & $ -341.2963243 $/10 & $ -527.5694173 $/10\\
revTPSS & $ -128.9242010 $/10 & $ -162.2273272 $/10 & $ -200.0077708 $/10 & $ -341.1618278 $/10 & $ -527.3782603 $/10\\
TASKCC & $ -128.9847733 $/10 & $ -162.2753046 $/10 & $ -200.0383120 $/15 & $ -341.1235642 $/10 & $ -527.3566109 $/10\\
MS0 & $ -128.9818356 $/15 & NoCBS & NoCBS & NoCBS & $ -527.5880852 $/20\\
rSCAN & $ -128.9723924 $/10 & $ -162.2983845 $/20 & $ -200.0958519 $/15 & $ -341.3288363 $/15 & $ -527.6283474 $/15\\
r$^2$SCAN & $ -128.9348395 $/10 & NoSCF & $ -200.0443066 $/15 & $ -341.2500476 $/15 & $ -527.5177200 $/15\\
r$^2$SCAN01 & $ -128.9394874 $/10 & $ -162.2600473 $/15 & $ -200.0501446 $/15 & $ -341.2582840 $/20 & $ -527.5287026 $/15\\
\hline
$\omega$B97X-noV & $ -128.9808707 $/5 & $ -162.2959918 $/5 & $ -200.0901465 $/5 & $ -341.2765096 $/5 & $ -527.5527095 $/10\\
B97M-noV & $ -128.9741484 $/10 & NoSCF & $ -200.0825748 $/15 & NoSCF & $ -527.4912402 $/10\\
$\omega$B97M-noV & $ -128.9939646 $/10 & NoSCF & $ -200.1215024 $/10 & NaN & $ -527.6026146 $/10\\
\hline
M08-HX & $ -128.9488829 $/10 & NoCBS & NoCBS & $ -341.2642177 $/15 & $ -527.5522818 $/20\\
MN12-SX & $ -128.9439150 $/10 & NoSCF & NoCBS & NoSCF & $ -527.5637752 $/15\\
MN12-L & $ -128.9511777 $/10 & NoSCF & NoSCF & $ -341.2968361 $/15 & $ -527.5498561 $/15\\
MN15 & $ -128.9582835 $/10 & NoCBS & $ -200.0789652 $/10 & $ -341.2689683 $/10 & $ -527.6036546 $/10\\
MN15-L & $ -128.9359083 $/10 & NoCBS & $ -200.0757004 $/15 & $ -341.2873822 $/15 & $ -527.5886820 $/10\\
revM06 & $ -128.9455051 $/10 & NoSCF & NoSCF & $ -341.2549085 $/10 & $ -527.5410881 $/10\\
revM06-L & $ -128.9535087 $/10 & NoSCF & $ -200.0648304 $/15 & $ -341.2576943 $/10 & $ -527.5364675 $/10\\
M06-SX & $ -128.9442104 $/10 & NoSCF & NoSCF & $ -341.2612515 $/10 & $ -527.5470863 $/10\\
revM11 & $ -128.9442958 $/10 & $ -162.2556667 $/10 & $ -200.0475869 $/10 & $ -341.2534754 $/10 & $ -527.5401522 $/10\\
\end{tabular}

\caption{Reference energies in $E_{h}$ with the finite element method / number
of radial elements required to reach the converged energy for the
Ne, Na, Mg, P, and Ar atoms. The notation is analogous to that in
\tabref{H-N}. \label{tab:Ne-Ar}}
\end{table*}

\begin{table}
\small
\begin{tabular}{ll}
Functional & Threshold \\
PW92 & $ 10^{-9} $\\
PBE & $ 10^{-10} $\\
BLYP & $ 10^{-11} $\\
B3LYP & $ 10^{-11} $\\
B97 & $ 10^{-11} $\\
TPSS & $ 10^{-10} $\\
revTPSS & $ 10^{-10} $\\
TASKCC & $ 10^{-10} $\\
MS0 & $ 10^{-10} $\\
rSCAN & $ 10^{-9} $\\
r$^2$SCAN & $ 10^{-10} $\\
r$^2$SCAN01 & $ 10^{-10} $\\
$\omega$B97X-noV & $ 10^{-10} $\\
B97M-noV & $ 10^{-10} $\\
$\omega$B97M-noV & $ 10^{-10} $\\
M08-HX & $ 10^{-10} $\\
MN12-SX & $ 10^{-11} $\\
MN12-L & $ 10^{-11} $\\
MN15 & $ 10^{-11} $\\
MN15-L & $ 10^{-11} $\\
revM06 & $ 10^{-10} $\\
revM06-L & $ 10^{-10} $\\
M06-SX & $ 10^{-10} $\\
revM11 & $ 10^{-10} $\\
\end{tabular}

\caption{Largest density threshold $\epsilon\in[10^{-8},10^{-9},\dots,10^{-15}]$
that reproduces the reference energies of \tabref{H-N,Ne-Ar} to $0.1\mu E_{h}$.
Calculations that failed were not included in the analysis. \label{tab:Density-thresholds}}
\end{table}

\subsection{Gaussian-Basis Truncation Errors \label{subsec:Gaussian-basis-Truncation-Errors}}

Furnished with the converged fully numerical reference energies, we
are able to determine truncation errors of Gaussian basis sets. This
part of the study is partly motivated by our recent work in \citeref{Schwalbe2022_JCP_174113},
where we observed unexpectedly large basis set truncation errors for
hydrogen with the M06-L,\citep{Zhao2006_JCP_194101} M11-L,\citep{Peverati2012_JPCL_117}
and B97M-noV functionals. We later found M06-L and M11-L to be ill-behaved,\citep{Lehtola2023__}
exhibiting large oscillations in the density Laplacian $\nabla^{2}n$
in the ground state of the hydrogen atom, which explains the large
differences in energies in fully numerical and Gaussian-basis calculations. 

The question of the accuracy of meta-GGA functional energies in Gaussian
basis sets has not been addressed in the literature so far to the
best of our knowledge. The recent study of \citet{Kraus2020_JCTC_5712}
studied basis set extrapolations with modern density functionals using
\HelFEM{}, but does not appear to comment on the functional dependence
of the accuracy in total energy.

We consider the aug-pc-4 basis set\citep{Jensen2002_JCP_3502,Jensen2002_JCP_7372,Jensen2002_JCP_9234,Jensen2003_JCP_2459,Jensen2004_JCP_3463,Jensen2007_JPCA_11198}
in its fully uncontracted form (un-aug-pc-4), and our recent augmented
hydrogenic Gaussian basis set (AHGBS-9).\citep{Lehtola2020_JCP_134108}
The un-aug-pc-4 basis set has been optimized for the BLYP functional,\citep{Jensen2002_JCP_7372,Jensen2004_JCP_3463,Jensen2007_JPCA_11198}
while the AHGBS-9 basis set and its polarized counterparts are constructed
by considerations on one-electron ions, only.\citep{Lehtola2020_JCP_134108}
The hydrogenic basis sets of \citeref{Lehtola2020_JCP_134108} are
large even-tempered basis sets aimed for benchmark accuracy calculations
on atoms and molecules. Note that even-tempered basis sets are often
used for studies on basis set completeness for their favorable properties.\citep{Kutzelnigg1994_IJQC_447,Cherkes2009_IJQC_2996,Shaw2020_IJQC_26264} 

The truncation errors for the un-aug-pc-4 basis set are shown in \tabref{uapc4},
and the errors for AHGBS-9 basis set are shown in \tabref{ahgbs}.
The Gaussian-basis energies were determined with the basis set truncation
and density screening thresholds $\epsilon=10^{-12}$ and $\tilde{\epsilon}=10^{-12}$,
respectively, and a (500, 974) quadrature grid. 

We observe that the truncation errors are strongly functional dependent,
which is not surprising given the analogous FEM data in \tabref{H-N, Ne-Ar}.
Interestingly, even though un-aug-pc-4 has been optimized for the
BLYP functional, it is often not the functional for which the smallest
truncation error is observed: the lowest truncation error is often
also observed for the PW92 or B3LYP functional. This is of course
not surprising, as optimality of the exponents for a given functional
does not prevent the truncation error for another functional with
the same basis set being smaller.

In the case of the AHGBS-9 basis set, we observe that the smallest
truncation errors are achieved for all atoms in case of HF calculations.
This can likely be attributed to the quadratic character of the HF
energy functional compared to the more complicated mathematical form
of DFAs.

\begin{sidewaystable*}
\small
\begin{tabular}{lllllllllll}
Functional & H & He & Li & Be & N & Ne & Na & Mg & P & Ar\\
\hline\hline
HF & $ 4.1 \times 10^{-7} $ & $ 4.8 \times 10^{-6} $ & $ 9.4 \times 10^{-7} $ & $ 1.5 \times 10^{-6} $ & $ 5.3 \times 10^{-6} $ & $ 1.8 \times 10^{-5} $ & $ 1.2 \times 10^{-5} $ & $ 1.3 \times 10^{-5} $ & $ 1.6 \times 10^{-5} $ & $ 2.3 \times 10^{-5} $\\
PW92 & $ 3.2 \times 10^{-7} $ & $ 3.9 \times 10^{-6} $ & $ 9.4 \times 10^{-7} $ & $ 1.5 \times 10^{-6} $ & $ 4.8 \times 10^{-6} $ & $ 1.6 \times 10^{-5} $ & $ 9.9 \times 10^{-6} $ & $ 1.1 \times 10^{-5} $ & $ 1.1 \times 10^{-5} $ & $ 1.8 \times 10^{-5} $\\
PBE & $ 5.3 \times 10^{-7} $ & $ 3.2 \times 10^{-6} $ & $ 1.1 \times 10^{-5} $ & $ 7.3 \times 10^{-6} $ & $ 9.4 \times 10^{-6} $ & $ 1.3 \times 10^{-5} $ & $ 5.9 \times 10^{-5} $ & $ 4.8 \times 10^{-5} $ & $ 3.1 \times 10^{-5} $ & $ 2.6 \times 10^{-5} $\\
BLYP & $ 3.4 \times 10^{-7} $ & $ 3.2 \times 10^{-6} $ & $ 2.3 \times 10^{-6} $ & $ 3.3 \times 10^{-6} $ & $ 5.5 \times 10^{-6} $ & $ 1.2 \times 10^{-5} $ & $ 1.1 \times 10^{-5} $ & $ 9.0 \times 10^{-6} $ & $ 1.4 \times 10^{-5} $ & $ 1.7 \times 10^{-5} $\\
B3LYP & $ 3.4 \times 10^{-7} $ & $ 3.3 \times 10^{-6} $ & $ 1.7 \times 10^{-6} $ & $ 2.5 \times 10^{-6} $ & $ 4.8 \times 10^{-6} $ & $ 1.3 \times 10^{-5} $ & $ 1.1 \times 10^{-5} $ & $ 8.9 \times 10^{-6} $ & $ 1.2 \times 10^{-5} $ & $ 1.5 \times 10^{-5} $\\
B97 & $ 2.4 \times 10^{-6} $ & $ 3.3 \times 10^{-6} $ & N/A & $ 1.2 \times 10^{-5} $ & N/A & $ 2.2 \times 10^{-5} $ & $ 1.8 \times 10^{-4} $ & $ 4.4 \times 10^{-5} $ & $ 1.2 \times 10^{-4} $ & $ 8.3 \times 10^{-5} $\\
\hline
TPSS & $ 4.0 \times 10^{-6} $ & $ 7.8 \times 10^{-6} $ & $ 9.1 \times 10^{-5} $ & $ 1.2 \times 10^{-4} $ & $ 1.2 \times 10^{-4} $ & $ 3.4 \times 10^{-5} $ & $ 1.9 \times 10^{-4} $ & $ 1.7 \times 10^{-4} $ & $ 2.1 \times 10^{-4} $ & $ 3.2 \times 10^{-4} $\\
revTPSS & $ 4.0 \times 10^{-6} $ & $ 5.6 \times 10^{-6} $ & $ 8.7 \times 10^{-5} $ & $ 1.1 \times 10^{-4} $ & $ 1.3 \times 10^{-4} $ & $ 3.9 \times 10^{-5} $ & $ 1.7 \times 10^{-4} $ & $ 1.7 \times 10^{-4} $ & $ 1.5 \times 10^{-4} $ & $ 1.7 \times 10^{-4} $\\
TASKCC & $ 3.7 \times 10^{-7} $ & $ 4.6 \times 10^{-6} $ & $ 1.9 \times 10^{-4} $ & $ 3.5 \times 10^{-4} $ & $ 5.1 \times 10^{-4} $ & $ 3.4 \times 10^{-4} $ & $ 5.4 \times 10^{-4} $ & $ 7.9 \times 10^{-4} $ & $ 8.5 \times 10^{-4} $ & $ 8.2 \times 10^{-4} $\\
MS0 & $ 3.3 \times 10^{-7} $ & $ 3.9 \times 10^{-6} $ & N/A & N/A & $ 2.4 \times 10^{-4} $ & $ 2.3 \times 10^{-4} $ & N/A & N/A & N/A & $ 9.4 \times 10^{-4} $\\
rSCAN & $ 3.6 \times 10^{-7} $ & $ 4.5 \times 10^{-6} $ & N/A & $ 5.7 \times 10^{-5} $ & $ 1.0 \times 10^{-4} $ & $ 5.8 \times 10^{-5} $ & $ 1.0 \times 10^{-4} $ & $ 1.4 \times 10^{-4} $ & $ 2.4 \times 10^{-4} $ & $ 1.8 \times 10^{-4} $\\
r$^2$SCAN & $ 3.7 \times 10^{-7} $ & $ 4.5 \times 10^{-6} $ & N/A & $ 6.5 \times 10^{-5} $ & $ 1.2 \times 10^{-4} $ & $ 8.3 \times 10^{-5} $ & N/A & $ 1.5 \times 10^{-4} $ & $ 2.8 \times 10^{-4} $ & $ 2.3 \times 10^{-4} $\\
r$^2$SCAN01 & $ 3.6 \times 10^{-7} $ & $ 4.5 \times 10^{-6} $ & $ 1.0 \times 10^{-4} $ & $ 6.2 \times 10^{-5} $ & $ 1.2 \times 10^{-4} $ & $ 8.2 \times 10^{-5} $ & $ 1.6 \times 10^{-4} $ & $ 1.5 \times 10^{-4} $ & $ 2.8 \times 10^{-4} $ & $ 2.3 \times 10^{-4} $\\
\hline
$\omega$B97X-noV & $ 4.4 \times 10^{-7} $ & $ 3.4 \times 10^{-6} $ & $ 2.2 \times 10^{-6} $ & $ 4.2 \times 10^{-6} $ & $ 6.6 \times 10^{-6} $ & $ 1.6 \times 10^{-5} $ & $ 1.8 \times 10^{-5} $ & $ 1.9 \times 10^{-5} $ & $ 2.1 \times 10^{-5} $ & $ 2.5 \times 10^{-5} $\\
B97M-noV & $ 1.5 \times 10^{-5} $ & $ 1.9 \times 10^{-5} $ & N/A & $ 8.4 \times 10^{-4} $ & N/A & $ 7.6 \times 10^{-5} $ & N/A & $ 5.5 \times 10^{-4} $ & N/A & $ 2.9 \times 10^{-4} $\\
$\omega$B97M-noV & $ 6.7 \times 10^{-6} $ & $ 7.1 \times 10^{-6} $ & N/A & $ 8.1 \times 10^{-5} $ & N/A & $ 2.5 \times 10^{-5} $ & N/A & $ 3.5 \times 10^{-4} $ & N/A & $ 1.3 \times 10^{-4} $\\
\hline
M08-HX & $ 3.6 \times 10^{-4} $ & $ 2.4 \times 10^{-3} $ & N/A & N/A & $ 2.3 \times 10^{-3} $ & $ 3.6 \times 10^{-3} $ & N/A & N/A & $ 5.3 \times 10^{-3} $ & $ 8.2 \times 10^{-3} $\\
MN12-SX & $ 1.5 \times 10^{-4} $ & $ 7.4 \times 10^{-4} $ & N/A & N/A & $ 2.9 \times 10^{-3} $ & $ 2.5 \times 10^{-3} $ & N/A & N/A & N/A & $ 5.6 \times 10^{-3} $\\
MN12-L & $ 6.8 \times 10^{-5} $ & $ 1.5 \times 10^{-3} $ & N/A & N/A & $ 2.5 \times 10^{-3} $ & $ 4.3 \times 10^{-3} $ & N/A & N/A & $ 6.5 \times 10^{-3} $ & $ 1.1 \times 10^{-2} $\\
MN15 & $ 1.7 \times 10^{-5} $ & $ 3.4 \times 10^{-5} $ & N/A & $ 2.7 \times 10^{-4} $ & $ 4.1 \times 10^{-4} $ & $ 2.4 \times 10^{-4} $ & N/A & $ 1.4 \times 10^{-3} $ & $ 2.0 \times 10^{-3} $ & $ 2.1 \times 10^{-3} $\\
MN15-L & $ 1.5 \times 10^{-5} $ & $ 4.4 \times 10^{-4} $ & N/A & N/A & $ 1.0 \times 10^{-3} $ & $ 9.0 \times 10^{-4} $ & N/A & $ 1.6 \times 10^{-3} $ & $ 2.3 \times 10^{-3} $ & $ 2.4 \times 10^{-3} $\\
revM06 & $ 9.7 \times 10^{-5} $ & $ 2.5 \times 10^{-4} $ & N/A & $ 3.2 \times 10^{-4} $ & $ 3.2 \times 10^{-4} $ & $ 4.6 \times 10^{-4} $ & N/A & N/A & $ 4.3 \times 10^{-4} $ & $ 5.7 \times 10^{-4} $\\
revM06-L & $ 1.6 \times 10^{-4} $ & $ 3.5 \times 10^{-4} $ & N/A & $ 3.6 \times 10^{-4} $ & $ 4.9 \times 10^{-4} $ & $ 6.8 \times 10^{-4} $ & N/A & $ 1.2 \times 10^{-3} $ & $ 1.1 \times 10^{-3} $ & $ 1.3 \times 10^{-3} $\\
M06-SX & $ 6.1 \times 10^{-5} $ & $ 1.7 \times 10^{-4} $ & N/A & $ 1.9 \times 10^{-4} $ & $ 1.9 \times 10^{-4} $ & $ 3.0 \times 10^{-4} $ & N/A & N/A & $ 2.3 \times 10^{-4} $ & $ 4.0 \times 10^{-4} $\\
revM11 & $ 6.1 \times 10^{-5} $ & $ 3.2 \times 10^{-5} $ & N/A & N/A & $ 3.5 \times 10^{-4} $ & $ 2.1 \times 10^{-4} $ & $ 2.7 \times 10^{-4} $ & $ 4.5 \times 10^{-4} $ & $ 8.6 \times 10^{-4} $ & $ 8.9 \times 10^{-4} $\\
\end{tabular}

\caption{Truncation errors in $E_{h}$ for the un-aug-pc-4 Gaussian basis set,
computed using the reference energies accurate to $10^{-7}E_{h}$
given in \tabref{H-N, Ne-Ar}. N/A: The data is not available as converged
fully numerical energies could not be determined. \label{tab:uapc4}}
\end{sidewaystable*}

\begin{sidewaystable*}
\small
\begin{tabular}{lllllllllll}
Functional & H & He & Li & Be & N & Ne & Na & Mg & P & Ar\\
\hline\hline
HF & $\sim 0$ & $\sim 0$ & $\sim 0$ & $\sim 0$ & $ 1.3 \times 10^{-7} $ & $ 3.3 \times 10^{-7} $ & $ 4.4 \times 10^{-7} $ & $ 5.7 \times 10^{-7} $ & $ 1.1 \times 10^{-6} $ & $ 2.1 \times 10^{-6} $\\
PW92 & $\sim 0$ & $\sim 0$ & $ 4.9 \times 10^{-8} $ & $ 5.7 \times 10^{-8} $ & $ 1.4 \times 10^{-7} $ & $ 3.3 \times 10^{-7} $ & $ 4.9 \times 10^{-7} $ & $ 6.8 \times 10^{-7} $ & $ 1.4 \times 10^{-6} $ & $ 2.6 \times 10^{-6} $\\
PBE & $\sim 0$ & $\sim 0$ & $ 1.5 \times 10^{-5} $ & $ 4.9 \times 10^{-6} $ & $ 2.8 \times 10^{-6} $ & $ 1.1 \times 10^{-6} $ & $ 1.1 \times 10^{-5} $ & $ 4.0 \times 10^{-6} $ & $ 1.8 \times 10^{-5} $ & $ 2.2 \times 10^{-5} $\\
BLYP & $\sim 0$ & $ 7.4 \times 10^{-8} $ & $ 3.3 \times 10^{-6} $ & $ 3.3 \times 10^{-6} $ & $ 1.2 \times 10^{-6} $ & $ 8.7 \times 10^{-7} $ & $ 1.6 \times 10^{-6} $ & $ 1.5 \times 10^{-6} $ & $ 8.2 \times 10^{-6} $ & $ 1.4 \times 10^{-5} $\\
B3LYP & $\sim 0$ & $ 6.1 \times 10^{-8} $ & $ 1.8 \times 10^{-6} $ & $ 1.9 \times 10^{-6} $ & $ 7.1 \times 10^{-7} $ & $ 6.3 \times 10^{-7} $ & $ 1.1 \times 10^{-6} $ & $ 1.2 \times 10^{-6} $ & $ 5.0 \times 10^{-6} $ & $ 8.4 \times 10^{-6} $\\
B97 & $ 1.1 \times 10^{-7} $ & $\sim 0$ & N/A & $ 9.5 \times 10^{-6} $ & N/A & $ 3.0 \times 10^{-6} $ & $ 2.3 \times 10^{-5} $ & $ 5.8 \times 10^{-6} $ & $ 8.5 \times 10^{-5} $ & $ 6.3 \times 10^{-5} $\\
\hline
TPSS & $ 1.4 \times 10^{-7} $ & $\sim 0$ & $ 9.0 \times 10^{-5} $ & $ 1.3 \times 10^{-4} $ & $ 6.8 \times 10^{-5} $ & $ 5.8 \times 10^{-6} $ & $ 3.5 \times 10^{-5} $ & $ 7.4 \times 10^{-5} $ & $ 1.3 \times 10^{-4} $ & $ 1.5 \times 10^{-4} $\\
revTPSS & $ 2.3 \times 10^{-7} $ & $ 1.2 \times 10^{-7} $ & $ 8.2 \times 10^{-5} $ & $ 1.1 \times 10^{-4} $ & $ 8.3 \times 10^{-5} $ & $ 7.8 \times 10^{-6} $ & $ 4.8 \times 10^{-5} $ & $ 1.0 \times 10^{-4} $ & $ 1.2 \times 10^{-4} $ & $ 1.2 \times 10^{-4} $\\
TASKCC & $\sim 0$ & $\sim 0$ & $ 2.9 \times 10^{-4} $ & $ 3.7 \times 10^{-4} $ & $ 2.1 \times 10^{-4} $ & $ 2.9 \times 10^{-5} $ & $ 2.7 \times 10^{-4} $ & $ 3.9 \times 10^{-4} $ & $ 6.7 \times 10^{-4} $ & $ 4.9 \times 10^{-4} $\\
MS0 & $\sim 0$ & $\sim 0$ & N/A & N/A & $ 2.0 \times 10^{-4} $ & $ 2.0 \times 10^{-4} $ & N/A & N/A & N/A & $ 7.6 \times 10^{-4} $\\
rSCAN & $\sim 0$ & $\sim 0$ & N/A & $ 4.9 \times 10^{-5} $ & $ 7.3 \times 10^{-5} $ & $ 1.8 \times 10^{-5} $ & $ 6.5 \times 10^{-5} $ & $ 6.7 \times 10^{-5} $ & $ 1.6 \times 10^{-4} $ & $ 1.3 \times 10^{-4} $\\
r$^2$SCAN & $\sim 0$ & $\sim 0$ & N/A & $ 5.4 \times 10^{-5} $ & $ 8.5 \times 10^{-5} $ & $ 2.7 \times 10^{-5} $ & N/A & $ 8.6 \times 10^{-5} $ & $ 2.1 \times 10^{-4} $ & $ 1.8 \times 10^{-4} $\\
r$^2$SCAN01 & $\sim 0$ & $\sim 0$ & $ 1.1 \times 10^{-4} $ & $ 5.2 \times 10^{-5} $ & $ 8.3 \times 10^{-5} $ & $ 2.7 \times 10^{-5} $ & $ 1.1 \times 10^{-4} $ & $ 8.5 \times 10^{-5} $ & $ 2.1 \times 10^{-4} $ & $ 1.8 \times 10^{-4} $\\
\hline
$\omega$B97X-noV & $\sim 0$ & $\sim 0$ & $ 2.6 \times 10^{-6} $ & $ 1.8 \times 10^{-6} $ & $ 2.3 \times 10^{-6} $ & $ 9.7 \times 10^{-7} $ & $ 2.8 \times 10^{-6} $ & $ 2.9 \times 10^{-6} $ & $ 7.4 \times 10^{-6} $ & $ 1.7 \times 10^{-5} $\\
B97M-noV & $ 5.2 \times 10^{-6} $ & $ 1.2 \times 10^{-7} $ & N/A & $ 8.5 \times 10^{-4} $ & N/A & $ 2.5 \times 10^{-5} $ & N/A & $ 2.9 \times 10^{-4} $ & N/A & $ 2.8 \times 10^{-4} $\\
$\omega$B97M-noV & $ 4.9 \times 10^{-7} $ & $ 1.0 \times 10^{-6} $ & N/A & $ 1.4 \times 10^{-4} $ & N/A & $ 3.8 \times 10^{-6} $ & N/A & $ 2.3 \times 10^{-4} $ & N/A & $ 7.0 \times 10^{-5} $\\
\hline
M08-HX & $ 3.2 \times 10^{-5} $ & $ 4.7 \times 10^{-4} $ & N/A & N/A & $ 1.4 \times 10^{-3} $ & $ 2.1 \times 10^{-3} $ & N/A & N/A & $ 3.8 \times 10^{-3} $ & $ 7.6 \times 10^{-3} $\\
MN12-SX & $ 1.6 \times 10^{-5} $ & $ 1.0 \times 10^{-4} $ & N/A & N/A & $ 1.3 \times 10^{-3} $ & $ 4.8 \times 10^{-4} $ & N/A & N/A & N/A & $ 3.7 \times 10^{-3} $\\
MN12-L & $ 1.3 \times 10^{-6} $ & $ 1.4 \times 10^{-4} $ & N/A & N/A & $ 1.1 \times 10^{-3} $ & $ 9.1 \times 10^{-4} $ & N/A & N/A & $ 3.7 \times 10^{-3} $ & $ 7.2 \times 10^{-3} $\\
MN15 & $ 9.3 \times 10^{-7} $ & $ 6.9 \times 10^{-7} $ & N/A & $ 2.5 \times 10^{-4} $ & $ 1.8 \times 10^{-4} $ & $ 3.5 \times 10^{-5} $ & N/A & $ 4.1 \times 10^{-4} $ & $ 1.2 \times 10^{-3} $ & $ 1.1 \times 10^{-3} $\\
MN15-L & $ 2.3 \times 10^{-6} $ & $ 3.3 \times 10^{-6} $ & N/A & N/A & $ 4.5 \times 10^{-4} $ & $ 1.2 \times 10^{-4} $ & N/A & $ 7.5 \times 10^{-4} $ & $ 1.9 \times 10^{-3} $ & $ 1.5 \times 10^{-3} $\\
revM06 & $ 7.1 \times 10^{-5} $ & $ 8.3 \times 10^{-5} $ & N/A & $ 3.1 \times 10^{-4} $ & $ 2.2 \times 10^{-4} $ & $ 1.6 \times 10^{-4} $ & N/A & N/A & $ 3.8 \times 10^{-4} $ & $ 4.5 \times 10^{-4} $\\
revM06-L & $ 3.5 \times 10^{-5} $ & $ 1.2 \times 10^{-5} $ & N/A & $ 4.1 \times 10^{-4} $ & $ 1.4 \times 10^{-4} $ & $ 8.4 \times 10^{-5} $ & N/A & $ 4.8 \times 10^{-4} $ & $ 7.6 \times 10^{-4} $ & $ 8.2 \times 10^{-4} $\\
M06-SX & $ 2.0 \times 10^{-5} $ & $ 1.0 \times 10^{-5} $ & N/A & $ 2.0 \times 10^{-4} $ & $ 8.0 \times 10^{-5} $ & $ 3.4 \times 10^{-5} $ & N/A & N/A & $ 1.7 \times 10^{-4} $ & $ 1.8 \times 10^{-4} $\\
revM11 & $ 5.3 \times 10^{-6} $ & $ 1.0 \times 10^{-6} $ & N/A & N/A & $ 1.7 \times 10^{-4} $ & $ 8.3 \times 10^{-5} $ & $ 1.5 \times 10^{-4} $ & $ 3.0 \times 10^{-4} $ & $ 5.8 \times 10^{-4} $ & $ 5.6 \times 10^{-4} $\\
\end{tabular}

\caption{Truncation errors in $E_{h}$ for the AHGBS-9 Gaussian basis set,
computed using the reference energies accurate to $10^{-7}E_{h}$
given in \tabref{H-N, Ne-Ar}. The notation is the same as in \tabref{uapc4};
however, cases where the difference between the Gaussian and finite
element results is smaller than the accuracy of the finite element
results are marked by $\sim0$. \label{tab:ahgbs}}
\end{sidewaystable*}

\subsubsection{Examination into Differences Between un-aug-pc-4 and AHGBS-9 \label{subsec:Examination-into-Differences}}

AHGBS-9 generally yields smaller basis set truncation errors than
un-aug-pc-4. This means AHGBS-9 is a better basis set, which is not
surprising given its large size: the basis sets of \citeref{Lehtola2020_JCP_134108}
were designed for high-accuracy calculations on small systems. 

Examining the truncation errors further, we observe that AHGBS-9 yields
a lower energy than un-aug-pc-4 in 191 calculations. However, we also
observe that the reverse is true in 12 calculations: un-aug-pc-4 yields
lower energies than AHGBS-9 in the PBE, BLYP, B3LYP, r$^{2}$SCAN01,
TASKCC, and $\omega$B97X-noV calculations on Li, and the TPSS, TASKCC,
B97M-noV, $\omega$B97M-noV, revM06-L, and M06-SX calculations on
Be. The observed issues therefore only affect Li and Be, for which
AHGBS-9 reproduces a lower total energy than un-aug-pc-4 in 16 calculations,
and a higher one in 12.

In the former case, the largest decreases in total energy going from
un-aug-pc-4 to AHGBS-9 are 15 $\mu E_{h}$ for Be with the MN15 functional,
and 4.9 $\mu E_{h}$ for Li with the revTPSS functional. In the latter
case, the largest decreases in total energy from AHGBS-9 to un-aug-pc-4
are 94 $\mu E_{h}$ for Li with the TASKCC functional, and 57 $\mu E_{h}$
for Be with the $\omega$B97M-noV functional.

We shall investigate these discrepancies in the results further by
examining the basis sets in detail. The examined configurations of
Li and Be only have $s$ electrons, and therefore examining the issue
reduces to examining the $s$ basis functions. Examining the $s$
exponents, we see that they span the range $5.93\times10^{-3}\ a_{0}^{-2}$
to $7.07\times10^{4}\ a_{0}^{-2}$ in un-aug-pc-4 for Li, while the
corresponding AHGBS-9 basis spans the range $7.05\times10^{-3}\ a_{0}^{-2}$
to $6.85\times10^{5}\ a_{0}^{-2}$, suggesting that AHGBS-9 may not
have sufficiently diffuse exponents in the case of Li. However, for
Be we observe exponents in the range $1.11\times10^{-2}\ a_{0}^{-2}$
to $1.39\times10^{5}\ a_{0}^{-2}$ in un-aug-pc-4, while in AHGBS-9
the range is $6.51\times10^{-3}\ a_{0}^{-2}$ to $1.22\times10^{6}\ a_{0}^{-2}$,
which thus fully covers the range of exponents included in un-aug-pc-4
and suggests that the issue could be a lack of completeness within
this range of exponents.

\begin{figure}
\begin{centering}
\includegraphics[width=1\linewidth]{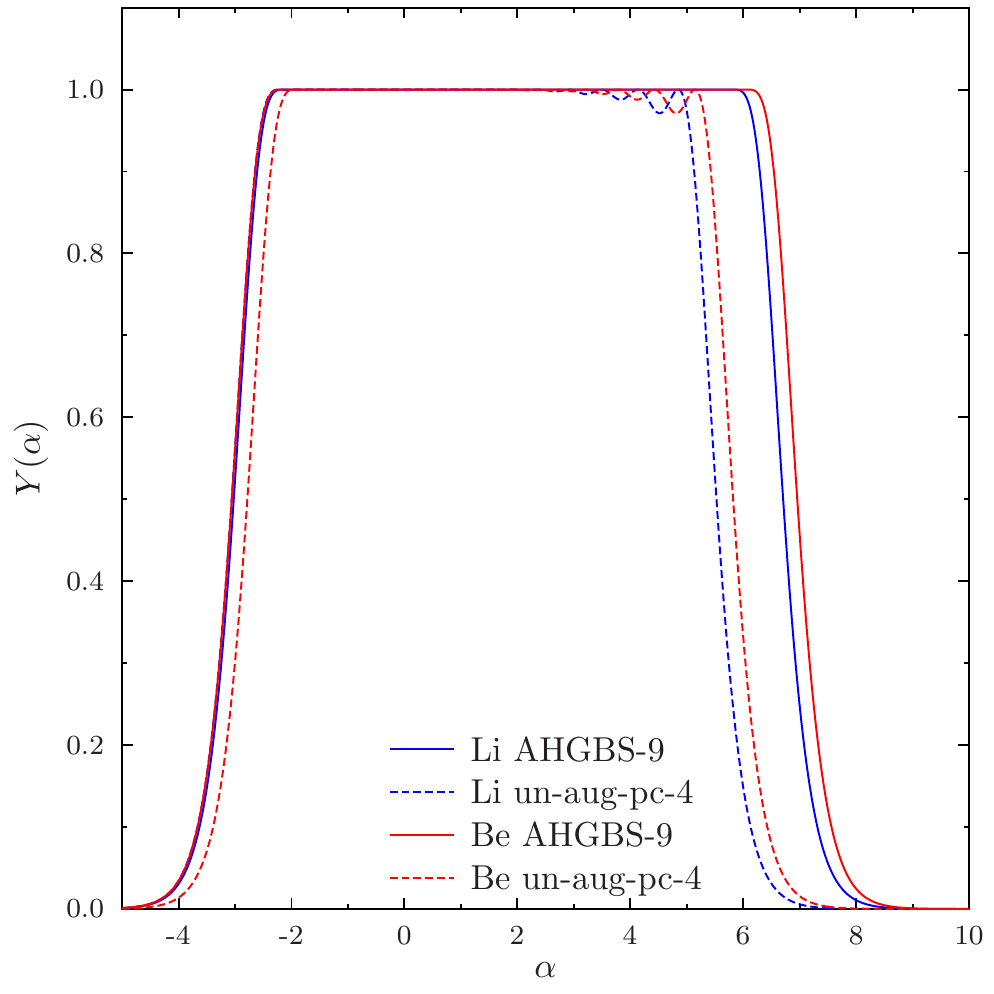}
\par\end{centering}
\caption{Completeness profile $Y(\alpha)$ for the $s$ exponents for Li and
Be in un-aug-pc-4 and AHGBS-9 basis sets. \label{fig:cpl}}

\end{figure}

The completeness profiles\citep{Chong1995_CJC_79}
\begin{equation}
Y(\alpha)=\sum_{\mu\nu}\langle\alpha|\mu\rangle\langle\mu|\nu\rangle^{-1}\langle\nu|\mu\rangle,\label{eq:cpl}
\end{equation}
where $\alpha$ is a test function, $\mu$ and $\nu$ are atomic basis
functions and $\langle\mu|\nu\rangle^{-1}$ denotes the $\mu,\nu$
element of the inverse overlap matrix offer a visual tool to inspect
the completeness of the studied basis sets. Gaussian functions with
exponents $\alpha$ can be expanded exactly in the basis if $Y(\alpha)=1$,
while functions that are orthogonal to the basis set have $Y(\alpha)=0$.
The completeness profiles for the studied $s$ functions in the Li
and Be basis sets are shown in \figref{cpl}. 

\begin{figure}
\begin{centering}
\includegraphics[width=1\linewidth]{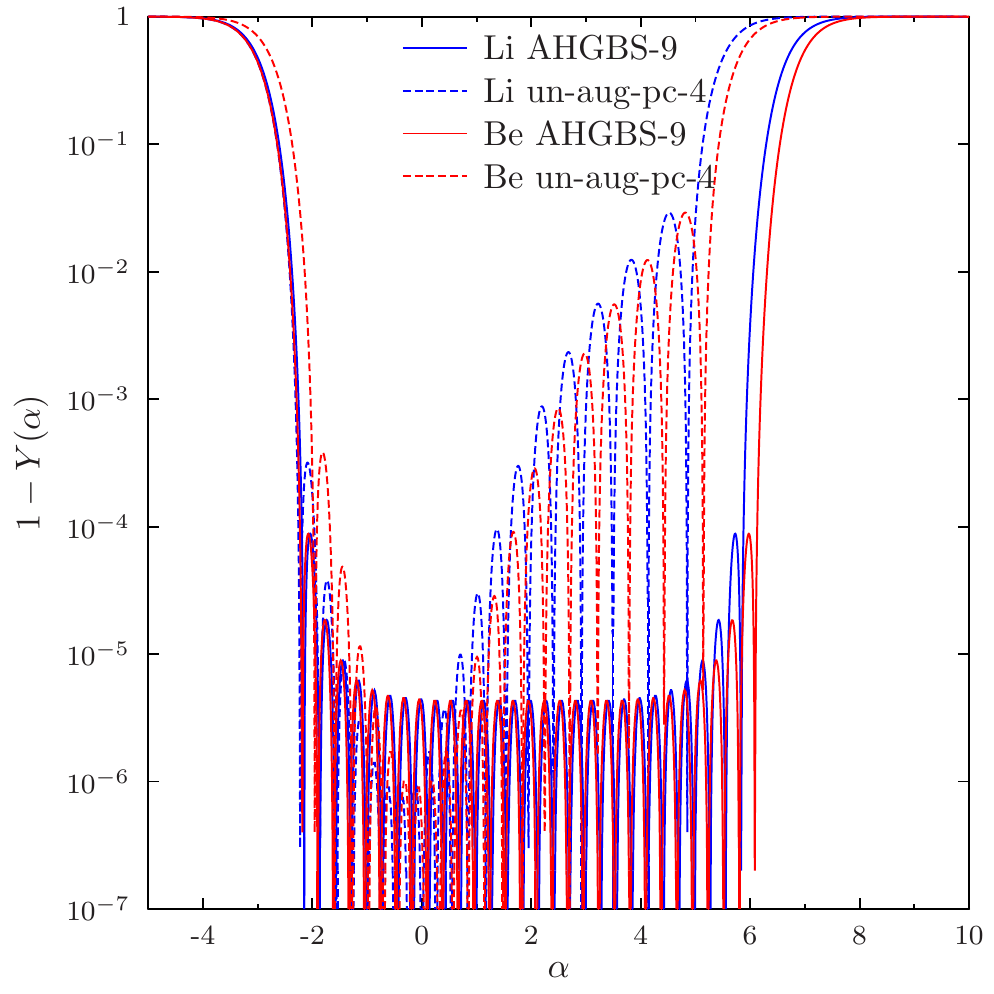}
\par\end{centering}
\caption{Complementary completeness profile $1-Y(\alpha)$ for the $s$ exponents
for Li and Be in un-aug-pc-4 and AHGBS-9 basis sets. Note use of a
logarithmic $y$ axis. \label{fig:complcpl}}
\end{figure}

\Figref{cpl} features oscillations in the profile of un-aug-pc-4
for large exponents $\alpha$, while the profile for the large AHGBS-9
basis set is flat. These features are even clearer when examining
the difference $1-Y(\alpha)$ in \figref{complcpl}: the AHGBS-9 basis
appears considerably more flexible than un-aug-pc-4 in the case of
Be in the same range of exponents.

We can thus summarize that a more complete Gaussian basis set may
reproduce a higher energy than a smaller, less complete Gaussian basis
set, depending on the functional. This indicates that the basis set
truncation errors observed for various combinations of Gaussian basis
sets and density functionals may have non-trivial dependence on the
actual exponent values. The dependence of the optimal exponents on
the DFA is caused by the differences in the optimal radial orbitals
of various functionals. Our data in \tabref{uapc4, ahgbs} demonstrate
that the use of a fixed Gaussian basis set can introduce functional
dependent errors in total energies in the range of tens to hundreds
of microhartree.

These findings again underline the importance of the present contribution
in introducing fully numerical methods for the reliable determination
of CBS limit energies. The issues discovered with the functional dependence
of the exponents also underline another part of our discussion. For
example, in the case of Be, using a converged PBE density to start
a r$^{2}$SCAN calculation shows that the PBE orbitals yield a total
energy which is 0.957 m$E_{h}$ higher than that for the converged
r$^{2}$SCAN orbitals. As we have discussed in \citeref{Lehtola2019_IJQC_25968},
NAOs are analogous to contracted basis sets, and this error is nothing
but the contraction error made when using a minimal NAO basis for
the PBE functional in a r$^{2}$SCAN calculation. Although additional
basis functions to allow breathing and polarization will allow for
energy lowerings, the error of the minimal basis will reintroduce
BSSE in calculations. We again underline that NAOs should be formed
with the same DFA used in a polyatomic calculation to eliminate errors
arising from differences in the optimal form of the radial orbitals. 

\subsection{Representative Timings \label{subsec:Representative-Timings}}

Exemplifying the discussion of \citeref{Lehtola2022_WIRCMS_1610}
on what kinds of science can be done on today's commodity hardware
with free and open source software, the rapidity of the present implementation
in \HelFEM{} is demonstrated by calculations on the author's laptop
running an Intel Core i5-1235U processor. For this demonstration,
we choose the Be, Ar, and Xe atoms, and compare the present symmetry
aware implementation with the general implementation of \citeref{Lehtola2019_IJQC_25945}.
As methods, we pick HF, PW92, PBE, B3LYP, TPSS, and r$^{2}$SCAN from
our previously used selection. 

The resulting energies and timings are shown in \tabref{Timings-for-calculations}.
As \HelFEM{} is a new project, the code has not been heavily optimized.
Although many things could be done to optimize its performance, the
code is usable in present form and fast enough to pursue investigations
into the numerical stability of density functional approximations,
for instance. 

As the timings were obtained on the same machine with largely the
same code, they give a good idea of the speedups achieved by the use
of symmetry. The most time per iteration in the general program is
spent on building the DFA components of the Fock matrix. This is also
a part that experiences major speedups due to the use of symmetry
to eliminate the angular degrees of freedom, as the quadrature over
the solid angle is not needed, nor is the handling and pointwise evaluation
of the spherical harmonics. The speedups are larger for DFT than for
HF, and increase going from LDAs to GGAs to meta-GGAs.

Although obtaining results for heavy atoms with the general program
may take up to tens of minutes, employing symmetry allows obtaining
CBS limit results on commodity hardware in a matter of seconds.

\begin{table*}
\begin{centering}
\subfloat[Be, 15 radial elements]{\begin{centering}
\begin{tabular}{llllll}
Method & $E_{\text{gen}}$ & $t_{\text{gen}}$ & $E_{\text{sym}}$ & $t_{\text{sym}}$ & Speedup\tabularnewline
\hline 
\hline 
HF & -14.5730232 & 0.4 & -14.5730232 & 0.3 & 1.5\tabularnewline
PW92 & -14.4464735 & 0.6 & -14.4464735 & 0.2 & 2.6\tabularnewline
PBE & -14.6299477 & 1.8 & -14.6299477 & 0.3 & 5.7\tabularnewline
B3LYP & -14.6733282 & 1.9 & -14.6733282 & 0.3 & 6.6\tabularnewline
TPSS & -14.6717170 & 2.6 & -14.6717170 & 0.3 & 7.5\tabularnewline
r$^{2}$SCAN & -14.6490866 & 2.5 & -14.6490866 & 0.4 & 6.7\tabularnewline
\end{tabular}
\par\end{centering}
}
\par\end{centering}
\begin{centering}
\subfloat[Ar, 15 radial elements]{\begin{centering}
\begin{tabular}{llllll}
Method & $E_{\text{gen}}$ & $t_{\text{gen}}$ & $E_{\text{sym}}$ & $t_{\text{sym}}$ & Speedup\tabularnewline
\hline 
\hline 
HF & -526.8175128 & 4.0 & -526.8175128 & 0.9 & 4.5\tabularnewline
PW92 & -525.9397934 & 8.2 & -525.9397934 & 0.6 & 14.7\tabularnewline
PBE & -527.3461288 & 14.2 & -527.3461288 & 0.7 & 20.1\tabularnewline
B3LYP & -527.5678350 & 12.3 & -527.5678350 & 0.8 & 15.0\tabularnewline
TPSS & -527.5694173 & 25.0 & -527.5694173 & 0.8 & 30.5\tabularnewline
r$^{2}$SCAN & -527.5177200 & 24.4 & -527.5177200 & 0.8 & 30.2\tabularnewline
\end{tabular}
\par\end{centering}
}
\par\end{centering}
\begin{centering}
\subfloat[Xe, 25 radial elements]{\centering{}%
\begin{tabular}{llllll}
Method & $E_{\text{gen}}$ & $t_{\text{gen}}$ & $E_{\text{sym}}$ & $t_{\text{sym}}$ & Speedup\tabularnewline
\hline 
\hline 
HF & -7232.1383639 & 183.1 & -7232.1383639 & 5.8 & 31.4\tabularnewline
PW92 & -7228.8341637 & 182.3 & -7228.8341637 & 5.1 & 35.7\tabularnewline
PBE & -7234.2332120 & 296.2 & -7234.2332120 & 7.3 & 40.6\tabularnewline
B3LYP & -7234.8674339 & 227.9 & -7234.8674339 & 6.1 & 37.5\tabularnewline
TPSS & -7234.4363678 & 420.1 & -7234.4363678 & 6.9 & 61.1\tabularnewline
r$^{2}$SCAN & -7234.8086847 & 471.9 & -7234.8086847 & 8.6 & 55.1\tabularnewline
\end{tabular}}
\par\end{centering}
\caption{Timings in seconds for various calculations on the Be, Ar, and Xe
atoms. The first and second columns show the energy reproduced $E_{\text{gen}}$
and the time $t_{\text{gen}}$ taken by the general program of \citeref{Lehtola2019_IJQC_25945}.
The third and fourth show the respective values $E_{\text{sym}}$
and $t_{\text{sym}}$ for the symmetry-aware program of this work
and \citeref{Lehtola2020_PRA_12516}. The last column shows the speedup
of using symmetry. All energies are in Hartree and times in seconds.
\label{tab:Timings-for-calculations}}
\end{table*}

\section{Summary and Discussion \label{sec:Summary-and-Discussion}}

We have presented the formalism necessary to implement meta-GGA functionals
in atomic calculations within the finite element method, and implemented
it in the free and open source \HelFEM{} program. Furnished with
the new implementation, we carried out a large number of calculations
with 31 density functionals on the 10 closed-shell or half-closed-shell
atoms from H to Ar to determine total energies converged to within
$0.1\mu E_{h}$ with respect to all parameters controlling the calculation:
the radial basis set, the quadrature scheme, as well as the density
threshold in the density functionals' implementation in Libxc\citep{Lehtola2018_S_1}.
Excluding the non-converging calculations, we found that a density
screening threshold of $10^{-11}\ a_{0}^{-3}$ was able to reproduce
total energies converged to 0.1 $\mu E_{h}$ for all studied functionals. 

Ill behavior was observed in several density functionals. The Li and
Na atoms proved to be the hardest systems in this study, which we
attribute to their extended electronic structure. Pathological behavior
was discussed for several functionals for the Li and Na atoms, where
the diagonalization of a good initial guess results in thousand-hartree
increases of the total energy. This points to issues with large derivatives,
which were not examined in our recent study on the numerical behavior
of density functionals,\citep{Lehtola2022_JCP_174114} and whose study
was one of the central motivations of this work, as fully numerical
calculations are stringent tests of density functionals' behavior.

Equipped with the fully numerical CBS limit energies, we proceeded
to study basis set truncation errors in the AHGBS-9\citep{Lehtola2020_JCP_134108}
and aug-pc-4\citep{Jensen2002_JCP_3502,Jensen2002_JCP_7372,Jensen2002_JCP_9234,Jensen2003_JCP_2459,Jensen2004_JCP_3463,Jensen2007_JPCA_11198}
basis sets in fully uncontracted form (un-aug-pc-4). The truncation
errors were found to be strongly dependent on the functional. Although
AHGBS-9 is designed for benchmark studies and is thereby much larger
than un-aug-pc-4, we found that un-aug-pc-4 afforded a lower total
energy than AHGBS-9 in 12 out of 28 calculations on Li and Be. (For
all other systems, AHGBS-9 yielded systematically lower total energies.)
Even though un-aug-pc-4 was found to have a more diffuse exponent
than AHGBS-9 for Li, in the case of Be the un-aug-pc-4 exponents were
found to be included in the range of the exponents for AHGBS-9 and
completeness profiles confirmed that AHGBS-9 is a more complete basis
set. We therefore concluded that the use of fixed Gaussian exponents
can introduce functional dependent errors in the range of tens to
hundreds of microhartrees.

Our results underline the importance of fully numerical studies of
novel density functionals. The timings presented in this work demonstrate
that with the use of symmetry, functionals can be swiftly characterized
by a fully numerical calculation. Furthermore, our implementation
is open source and is freely available online for anyone for any purpose.

This study is the cornerstone on the road to employing modern finite
element techniques for molecular calculations with NAOs. We hope to
pursue along the path marked in \citeref{Lehtola2019_IJQC_25968}
by introducing open source software for NAO calculations in upcoming
work. However, as most algorithms required by a NAO program can be
formulated independently of other technical choices made in the implementation,
our plan is to pursue a modular approach. As we have recently reviewed
in \citeref{Lehtola2022_WIRCMS_1610}, standard, reusable open source
libraries like Libxc\citep{Lehtola2018_S_1} promote peer review,
the free exchange of ideas, and maintainability of software, and we
are convinced that such libraries merit more attention.

\section*{Acknowledgments}

I thank Fabien Tran for comments on the manuscript. We thank the National
Science Foundation for financial support under grant no. CHE-2136142,
and the Academy of Finland for financial support under project numbers
350282 and 353749. We thank CSC -- IT Centre for Science (Espoo,
Finland) for computational resources.

\clearpage{}

\bibliography{citations}

\begin{mcitethebibliography}{108}
\bibitem[Not()]{Note-2}
To simplify the discussion, we refer to Hartree--Fock as a density functional
  that lacks a semilocal energy expression of the form of \eqref{metaGGA-f} and
  that instead relies on 100\% exact exchange.\relax
\mciteBstWouldAddEndPunctfalse
\mciteSetBstMidEndSepPunct{\mcitedefaultmidpunct}
{}{\mcitedefaultseppunct}\relax
\EndOfBibitem
\end{mcitethebibliography}

\begin{tocentry}
\includegraphics{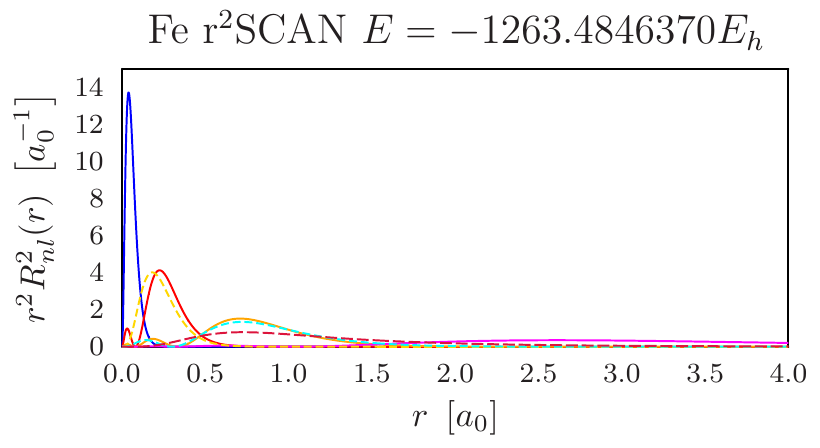}
\end{tocentry}

\end{document}